\documentclass[12pt]{article}
\pdfoutput=1

\usepackage{physics}
\usepackage{booktabs}
\usepackage{amssymb}
\usepackage{mwe}
\usepackage{amsmath}
\usepackage{graphicx}
\usepackage{caption}
\usepackage{subcaption}
\usepackage{tensor}
\usepackage{dsfont}
\usepackage{tikz}
\usepackage{tikz-feynman}
\usepackage{setspace}
\usepackage{titling} % for abstract
\usepackage{lipsum} % abstract filler
\usepackage[letterpaper, total={6.5in,8in}]{geometry}%,total={6in,8in}]{geometry}
\usepackage{hyperref}
\usepackage{authblk}
\usepackage{upgreek}
\usepackage{color}
\usepackage[normalem]{ulem}
\hypersetup{
    colorlinks=true,
    linkcolor=blue,
    filecolor=magenta,      
    urlcolor=cyan,
citecolor=cyan,
    pdftitle={Moving black holes in Horava gravity}
    }

\def\sectionautorefname~#1\null{Section (#1)\null}
\def\subsectionautorefname~#1\null{Section (#1)\null}
\def\equationautorefname~#1\null{(#1)\null}
\def\figureautorefname~#1\null{(Figure #1)\null}

\newcommand{\prd}[0]{\partial}

\def\ba{\begin{eqnarray}}
\def\ea{\end{eqnarray}}
\def\be{\begin{equation}}
\def\ee{\end{equation}}

\def\d{\partial}

\renewcommand{\l}{\lambda}

\renewcommand{\b}{\beta}

\newcommand{\g}{\gamma}

\newcommand{\vf}{\varphi}

\renewcommand{\a}{\alpha}
\renewcommand{\b}{\beta}
\renewcommand{\g}{\gamma}
\newcommand{\s}{\sigma}

\newcommand\bseq{\begin{subequations}}
\newcommand\eseq{\end{subequations}}

\numberwithin{equation}{section}

\allowdisplaybreaks[4]

\title{{\bf \Large Slowly moving black holes in khrono-metric model}}
\author[1]{A. Kovachik\footnote{kovachia@mcmaster.ca}}
\author[1,2]{S. Sibiryakov\footnote{ssibiryakov@perimeterinstitute.ca}}
\affil[1]{Department of Physics and Astronomy, McMaster University, 
Hamilton, ON L8S 4M1, Canada}
\affil[2]{Perimeter Institute for Theoretical Physics, Waterloo, Ontario, N2L 2Y5, Canada}
\begin{document}
%\begin{titlepage}
%\clearpage
\maketitle{}
\begin{abstract}
    We search for solutions describing slowly moving non-rotating black holes in the
    khrono-metric model, a modified gravity theory with preferred
   time (khronon) which arises at low energies from the
   non-projectable Ho\v rava gravity. We work in the decoupling limit when the
    back-reaction of the khronon on the metric is small and can be
    treated perturbatively. For a given black hole velocity, 
we find a family of solutions
    parameterized by the khronon propagation speed and regular
    everywhere outside the universal horizon. On the universal horizon
    they have a weak singularity manifesting itself in 
   a non-analyticity of the khronon field. Using the behavior of
   khronon at infinity we extract the leading black hole sensitivity
   for which we obtain a simple analytic expression valid throughout
   the phenomenologically allowed parameter space.

\end{abstract}

\thispagestyle{empty}

\newpage

{\hypersetup{hidelinks}
\tableofcontents
}

\section{Introduction}

Modified gravity theories with violation of Lorentz invariance have
received significant attention in the literature
\cite{Liberati:2013xla,Blas:2014aca}. Motivation often comes from 
quantum gravity. A concrete example is provided by the 
Ho\v rava's proposal \cite{horava_quantum_2009} to make gravity perturbatively
renormalizable by imposing invariance under
anisotropic scaling of time and space at high energies, see
\cite{Barvinsky:2023mrv,Herrero-Valea:2023zex}
for recent reviews. A so-called
projectable version of Ho\v rava gravity has indeed been proven
to be renormalizable \cite{Barvinsky:2015kil,Barvinsky:2017zlx} and
even asymptotically free
\cite{Barvinsky:2017kob,Barvinsky:2019rwn,Barvinsky:2021ubv,Barvinsky:2023uir,Barvinsky:2024svc}.
However, it suffers from an instability at low energies
\cite{Koyama:2009hc,blas_models_2011} and it 
is not clear at the moment if and how it can reproduce the observed
gravitational phenomenology. Partial results in favor of
renormalizability \cite{Bellorin:2022qeu,Bellorin:2022efu} exist also
for non-projectable Ho\v rava gravity which avoids the instability problem  \cite{Blas:2009qj}. At low energies it reduces to  Einstein's general
relativity (GR) coupled to a scalar field $\vf$ called khronon which
describes a preferred time coordinate
\cite{Germani:2009yt,Blas:2009yd,blas_models_2011,Wang:2017brl}. Geometrically, khronon
labels a foliation of spacetime by a family of spacelike
surfaces. This khrono-metric model is
parameterized by three dimensionless coupling constants conventionally denoted as
$\a$, $\b$, $\l$ and under certain
constraints on them passes all observational tests
\cite{EmirGumrukcuoglu:2017cfa}.    

The khrono-metric model is closely related
to the Einstein--aether theory
\cite{Jacobson:2000xp,Jacobson:2007veq,Oost:2018tcv}. In the latter
the preferred frame effects are described 
by a unit timelike vector $u^\mu$ called aether.
Imposing on aether the constraint of vanishing curl renders it
orthogonal to a spacelike foliation which obeys the khronon dynamics
\cite{Jacobson:2013xta,Bhattacharyya:2015uxt}. 

The khrono-metric model admits static spherically symmetric solutions
analogous to Schwarzschild black holes 
\cite{barausse_black_2011,blas_horava_2011}. 
Remarkably, they possess
non-trivial causal structure, despite the fact that breaking of
Lorentz invariance allows arbitrarily fast signal propagation in the khronon
rest-frame \cite{Bhattacharyya:2015gwa}. It happens that the leaves of khronon foliation extending
to infinity do not penetrate into the inner region of the black hole,
implying that no signal, no matter how fast, can escape from this region to
infinity. The boundary of this region has been called the universal
horizon. Several studies have demonstrated formation of such
black holes in spherical collapse \cite{Saravani:2013kva,Bhattacharyya:2015uxt,Bhattacharjee:2018nus,Franchini:2021bpt}. 
Rotating black hole solutions in the khrono-metric and Einstein-aether models have also been obtained: analytically in the limit of slow rotation \cite{Barausse:2012qh,Barausse:2015frm}, and numerically for general spin \cite{Adam:2021vsk}.

The solutions mentioned above represent black holes at rest in the reference frame described by the khronon. 
Realistic black holes, however, will move with respect to the khronon frame; for example, this will be the case for a black hole in a binary system.
 Due to breaking of Lorentz invariance, such black holes cannot be obtained from the static ones by a simple boost. The effect of khronon on moving compact bodies is parameterized by the so-called {\it sensitivities} which encode the dependence of the inertial mass of the body on its velocity with respect to the preferred frame \cite{foster_strong_2007}. The sensitivities play a key role in the binary system dynamics \cite{Will:2018ont}. In particular, the difference in sensitivities of the two companions leads to enhanced energy losses through 
dipolar gravitational radiation \cite{foster_strong_2007}. Refs.~\cite{yagi_constraints_2014,Gupta:2021vdj} calculated the sensitivities of neutron stars in khrono-metric and Einstein-aether models and used them to put strong constraints on the models from pulsar timing. Extending the calculation of sensitivities to black holes is strongly motivated by the rapid accumulation of gravity wave data from compact binaries \cite{KAGRA:2013rdx}, as well as prospects of future precision measurements of the gravitational radiation waveforms from extreme mass ratio inspirals \cite{Berry:2019wgg}.   

In general, finding a moving black hole solution presents a challenging task which requires solving a system of coupled partial differential equations. 
However, if the relative velocity $\mathbf{v}$ is small, ${\rm v}\ll 1$,
the moving solution is expected to deviate little from the static one
and the correction can be searched for as a perturbation linear in
$\mathbf{v}$. 
This strategy was adopted in
Ref.~\cite{ramos_constraints_2019} which searched for moving (non-rotating) black holes in the khrono-metric model. Their numerical analysis, however, did not yield any regular solutions at general values of the khronon couplings $\a,\b,\l$, failing to satisfy simultaneously the conditions of asymptotic flatness at infinity and regularity at the matter and spin-0 horizons.  
The former represents the causal horizon for matter
propagating with the speed of light, whereas the latter is determined
by the velocity of the low-energy khronon waves and is also referred
to as the khronon ``sound'' horizon. Both these horizons lie outside
the universal horizon and thus a singularity at them would be
accessible to an asymptotic observer using fast enough probes.

Ref.~\cite{ramos_constraints_2019} could find solutions 
satisfying all regularity conditions only if
two of the khronon couplings were set to be exactly zero, $\a=\b=0$.
The metric in this case exactly
coincides with the Schwarzschild metric and hence is regular down to
the central singularity, whereas the khronon becomes a
``stealth'' field with vanishing energy-momentum tensor (EMT)
\cite{Berglund:2012bu,Franchini:2021bpt}.
The black hole sensitivities then vanish identically, and the khronon field has no effect on the binary motion.

Absence of any regular moving black holes for non-zero $\a$ or $\b$ appears surprising, given that consistency of the khrono-metric model as an effective low-energy theory requires $\a>0$ \cite{blas_models_2011}, and in general all three couplings $\a$, $\b$, $\l$ are expected to be non-vanishing.\footnote{This does not exclude a possible hierarchy between these couplings which is imposed by the observational constraints, see Sec.~\ref{ssec:constraints}.}  
This motivates us to revisit 
the existence of slowly moving black holes in
the khrono-metric model at non-zero values of all three couplings
$\a$, $\b$, $\l$. 
Rather than considering the full set of metric--khronon equations as was done in \cite{ramos_constraints_2019}, we simplify the problem by exploiting the so-called decoupling limit \cite{blas_models_2011}. 
Namely, we assume all parameters $\a,\b,\l$ to be much less than unity (albeit non-zero). As reviewed in Sec.~\ref{sec:notationbkg}, this is in accordance with the existing observational constraints. 
This allows us to treat the effect of the khronon
EMT on the metric perturbatively. In the leading order it can be
neglected altogether and the  
problem reduces to finding the
khronon configuration in an external Schwarzschild metric. 
This strategy was applied to 
static khrono-metric black holes in \cite{blas_horava_2011}. 
It simplifies the treatment of various horizons corresponding to 
the singular points of the khronon equation and makes analysis of the regularity conditions for the solutions at these points transparent. 

Contrary to Ref.~\cite{ramos_constraints_2019}, we find regular
khronon solutions for general values of $\a$, $\b$ and $\l$. The
solutions are analytic everywhere outside the universal horizon and
have a mild non-analyticity on the latter. The khronon EMT remains
bounded and small all the way down to the universal horizon justifying
the use of the decoupling limit. We conclude that our solutions
provide a valid leading-order approximation to moving black holes in
the khrono-metric model and their existence does not impose any
constraints on the parameters of the theory. In the limit $\a=\b=0$
we recover the stealth khronon configurations found in
\cite{ramos_constraints_2019}. 
The origin of discrepancy between our results and those of \cite{ramos_constraints_2019} in the case of general khronon couplings is left for future investigation.

We use our moving black hole solutions to extract the first black hole sensitivity parameter $\s$. Previous computations of sensitivities make use of the
modification of the metric far away from the moving body due to the
presence of the fields --- khronon or aether --- describing the
preferred frame. In this work we develop a simpler method which does
not require computing the perturbation of the metric and extracts the 
sensitivities from the long-distance
asymptotics of the 
khronon alone. This gives us an expression for $\sigma$ in terms of the khronon couplings and the
subleading coefficient in the khronon expansion at infinity. We show that the latter can
be well approximated by a simple analytic formula throughout the
whole phenomenologically allowed parameter space. The resulting expression for the sensitivity is independent of the black hole mass. 
Taking into account the existing constraints on the model parameters, the sensitivity happens to be rather small. Nevertheless, it represents an important ingredient for uncovering the effect of khronon on the gravitational waveforms.

The paper is organized as follows. In Sec.~\ref{sec:notationbkg} we
give an overview of the khrono-metric model. We describe its action and introduce the
decoupling limit in Sec.~\ref{ssec:decoupl}, whereas in Sec.~\ref{ssec:constraints} we summarize the observational constraints on the model parameters.
In Sec.~\ref{sec:stationary} we review the solutions
describing static
spherically symmetric black
holes. In Sec.~\ref{sec:movingbh} we turn to slowly moving black
holes. In Sec.~\ref{ssec:movingEOM} we derive equations for
perturbation of the khronon in the rest frame of the black hole. In
Sec.~\ref{sec:moving} we perform the analysis of its singular points
and formulate the conditions that a khronon perturbation must satisfy
in order to give a regular foliation. Numerical solutions with
required boundary conditions are constructed in
Sec.~\ref{sec:solutions}. Section~\ref{sec:infcchi} discusses
simplifications occurring in the limit of infinite khronon speed and
obtains the solution in this case. 
In Sec.~\ref{sec:sens} we find the khronon field by replacing the
black hole with an effective point particle and match it to the
asymptotics of the full numerical solution. This gives us the
black hole 
sensitivities. Section~\ref{sec:conc} is devoted to
discussion of our results and outlook. Several appendices contain auxiliary material.

\section{Khrono-metric model}\label{sec:notationbkg}

\subsection{Action and decoupling limit}
\label{ssec:decoupl}

We consider GR coupled to a scalar field $\vf$ called khronon. The
surfaces $\vf={\rm const}$ define a preferred foliation. We impose
that the gradient of $\vf$ is timelike and non-vanishing, so that
$\vf$ can be thought of as a preferred time coordinate. We further
impose an invariance of the theory under relabeling of the foliation
leaves, equivalent to the transformations
\be
\label{reparam}
\vf\mapsto\tilde \vf =f(\vf)\;,
\ee 
with $f$ being an arbitrary increasing function. 
An object with the smallest number of derivatives invariant under
these transformations is the normalized khronon gradient,\footnote{We
  use the $(+,-,-,-)$ signature 
  of the metric and set the speed of light to unity.}
\be
\label{umuphi}
u_\mu=X^{-1/2} \d_\mu\vf~,~~~~~X=g^{\mu\nu}\d_\mu\vf\d_\nu\vf\;.
\ee 
Below we will often refer to the vector $u_\mu$ as aether. Since it
has unit norm, we need to add further derivatives to obtain
non-trivial terms in the Lagrangian. Then the most general action in
the lowest order of derivatives reads \cite{blas_models_2011},
\begin{equation}\label{eqn:action} 
    S = -\frac{M^2}{2} \int \text{d}^4x \sqrt{-g}\Big[
    R +
    \alpha\, a_\mu a^\mu +
    \beta(\nabla_\mu u_\nu) (\nabla^\nu u^\mu) +
    \lambda\,\vartheta^2\Big]\;,
\end{equation}
where $R$ is the Ricci curvature, whereas 
$a_\mu=u^\nu\nabla_\nu u_\mu$ and $\vartheta=\nabla_\mu u^\mu$ are the 
aether acceleration and divergence, respectively. The three
dimensionless couplings $\a,\b,\l$ represent the free parameters of
the model. The mass 
scale $M$ is close to the Planck mass and is related to the Newton's
gravitational constant as
\be
\label{GNewt}
G_N=\frac{1}{8\pi M^2(1-\a/2)}\;.
\ee
Note that the action
(\ref{eqn:action}) contains up to four derivatives when written
explicitly in terms of the khronon field $\vf$. 
This, however, does not
lead to the Ostrogradski ghosts since one can show that only two time derivatives
survive in the preferred frame \cite{blas_models_2011}. The
khrono-metric model (\ref{eqn:action})
arises as the low-energy limit of non-projectable
Ho\v rava gravity \cite{horava_quantum_2009,Blas:2009qj}.

For our purposes, it is convenient to cast the action
(\ref{eqn:action}) into a slightly different form. Since aether is
hypersurface-orthogonal, it has vanishing curl,
\begin{equation}
\label{zerocurl}
\omega_\mu \equiv    \epsilon^{\mu\nu\rho\sigma}u_{\nu}\nabla_\rho
u_\sigma = 0\;. 
\end{equation}
Squaring this identity, we find
\begin{equation}
\label{a2umunu}
    a_\mu a^\mu - \frac{1}{2} u_{\mu\nu}u^{\mu\nu}=0\;,
\end{equation}
where we have introduced 
$
%\label{umunu}
u_{\mu\nu} \equiv \partial_\mu u_\nu - \partial_\nu u_\mu
$.
Using this relation and integrating by parts in the $\b$-term, we
arrive at
\begin{equation}\label{eqn:khrononaction}
    S = - \frac{M^2}{2}
\int \dd[4]{x} \sqrt{-g} \left[R+ \frac{\a}{2} 
u_{\mu\nu}u^{\mu\nu} + (\b+\l)\,\vartheta^2 
-\b R_{\mu\nu}u^\mu u^\nu\right].
\end{equation}
The equation of motion for the khronon has the form of a current
conservation,
\be
\label{khrononeqm}
\nabla_\nu J^\nu=0~,~~~~~
J^\nu= X^{-1/2} P_\mu^\nu \big[\a \nabla_\l u^{\mu\l}
-(\b+\l) \nabla^\mu \vartheta-\beta R^{\mu\l} u_\l\big]\;,
\ee 
where
$
%\label{proj}
P_\mu^\nu=\delta_\mu^\nu-u^\nu u_\mu
$
is a projector on the leaves of the foliation. Around a fixed background
with Minkowski metric $\eta_{\mu\nu}$ and khronon identified with the
time coordinate, $\bar\vf=t$, the linearized equations for the khronon
perturbations $\chi=\vf-\bar\vf$ takes the form,
\be
\label{waveeq}
\Delta \big[\a\,\ddot\chi-(\beta+\l)\Delta\chi\big]=0\;,
\ee
where $\Delta=\d_i\d_i$ is the spatial Laplacian. We see that the
khronon waves propagate with the speed
\be
\label{cchi}
c_\chi=\sqrt{\frac{\b+\l}{\a}}\;,
\ee
to which we will loosely refer as the khronon ``sound'' speed. It can
be smaller or bigger than unity, corresponding to subluminal and
superluminal khronon propagation, respectively.

The khronon EMT following from the action (\ref{eqn:khrononaction})
has the form,    
\begin{align}
T_{\mu\nu}=&\a M^2 \left[-u_{\mu\l}u_\nu^{~\l}
-(\nabla_\l u^{\l\rho}) u_\rho u_\mu u_\nu
+\frac{1}{4}g_{\mu\nu}u_{\l\rho} u^{\l\rho}\right]\notag\\
&+(\b+\l)M^2\left[-(g_{\mu\nu}+u_\mu u_\nu) u^\l\nabla_\l\vartheta
+u_\mu \nabla_\nu \vartheta +u_\nu \nabla_\mu \vartheta
-\frac{1}{2} g_{\mu\nu}\vartheta^2\right]\notag\\
&+\b M^2\bigg[
-\frac{1}{2} \nabla_\l\nabla_\mu(u^\l u_\nu)
-\frac{1}{2} \nabla_\l\nabla_\nu(u^\l u_\mu)
+\frac{1}{2} \nabla_\l\nabla^\l (u_\mu u_\nu)
+\frac{1}{2} g_{\mu\nu}\nabla_\l\nabla_\rho (u^\l u^\rho)
\notag\\
&\qquad\qquad
+R_{\l\mu}u^\l u_\nu+R_{\l\nu}u^\l u_\mu
-R^{\l\rho}u_\l u_\rho u_\mu u_\nu
-\frac{1}{2}g_{\mu\nu} R_{\l\rho} u^\l u^\rho\bigg]\;.
\label{khEMT}
\end{align}
Clearly, it is proportional to the khronon parameters $\a,\b,\l$ and is
suppressed when these parameters are small. 

This allows us to employ
the following strategy when looking for solutions of the khrono-metric
theory. In the leading order we can neglect the effect of the khronon
on the metric and take the latter to be a solution of GR. The problem
then amounts to finding an embedding of khronon into the fixed
metric which greatly simplifies the technical task. Instead of solving a coupled system of differential equations for the ten metric components and the khronon, with extra complications associated to the diffeomorphism invariance, we just need to solve a single equation for khronon in a fixed coordinate frame. This makes the analysis of the system and the properties of the solution transparent.  
We will refer to this regime of negligible back-reaction of khronon on the metric as ``decoupling limit'' and will
adopt it in what follows. As long as the obtained khronon field
is regular, its EMT will be small and its back-reaction on the metric
can be taken into account perturbatively. 

\subsection{Observational constraints}
\label{ssec:constraints}

Besides technical simplifications, taking $\a,\b,\l\ll 1$ is
motivated by the observational constraints on the khrono-metric model 
\cite{EmirGumrukcuoglu:2017cfa}. The detection of the
gravitational wave event GW170817 accompanied by the gamma-ray burst
GRB170817A
produced by a binary
neutron star merger 
strongly restricts the difference between
the speeds of the gravitational and electromagnetic waves \cite{LIGOScientific:2017zic}. 
This leads to a very strong bound:
\be
\label{betabound}
|\beta|\lesssim 10^{-15}\;.
\ee
Further constraints come from the measurements in the Solar
System which restrict post-Newtonian parameters characterizing the preferred frame effects in a weak
gravitational field: 
$
%\label{PPNbounds}
|\a_1^{\rm PPN}|\lesssim  10^{-4}$,
$
|\a_2^{\rm PPN}|\lesssim  10^{-7}
$
\cite{Will:2005va}.
In the decoupling limit, they are expressed through the model parameters
as \cite{blas_models_2011},
\be
\label{PPNformulas}
\a_1^{\rm PPN}=-4(\a-2\b)~,~~~~~
\a_2^{\rm PPN}=\frac{(\a-2\b)(\a-3\b-\l)}{2(\b+\l)}\;.
\ee
Combined with (\ref{betabound}), this implies
\be
\label{PPNcons}
|\a|\lesssim 2.5\cdot 10^{-5}~~\text{and}~~\l\approx \a\;,
\qquad \text{or}\qquad 
|\a|\lesssim 2\cdot 10^{-7}~~\text{and}~~\l\gg \a\;.
\ee
A tighter constraint on the post-Newtonian parameter $\a_1^{\rm PPN}$ was obtained from the observation of the triple system PSR J0337+1715 consisting of an inner pulsar -- white dwarf binary and another white dwarf in an outer orbit. Absence of differential acceleration between the members of the inner system in the gravitational field of the outer white dwarf constrains~\cite{Gupta:2021vdj}
\be
\label{alpha1PT}
|\a_1^{\rm PPN}|<2.4\cdot 10^{-5}~~~~\text{at}~~ 95\%~\text{CL}
\ee
yielding 
\be
\label{alphaPT}
|\a|<0.6\cdot 10^{-5}\;.
\ee
It has also been suggested that very tight constraints on the Lorentz violating post-Newtonian parameter $\a_2^{\rm PPN}$ can be obtained from timing of solitary pulsars \cite{Shao:2013wga} (see also \cite{Will:2014kxa}). 
However, translating them into the bounds on the model
  parameters requires treatment of strong gravitational field effects and modeling of the neutron star structure, so we do not consider them in detail. 
Finally, a constraint on $\l$ comes from cosmology \cite{Frusciante:2020gkx},\footnote{Interestingly, the analysis of \cite{Wen:2023wes} suggests that the cosmological data prefer a non-zero negative value of $\l\simeq-(6\pm 3)\cdot 10^{-3}$. This, however, together with the constraint (\ref{betabound}), contradicts the second stability condition in (\ref{stab}).}
\be
\label{lambdabound}
|\l|\lesssim 2\cdot 10^{-3}\;.
\ee
It is weaker than the bounds on $\a$ and
$\beta$. 

\begin{figure}[t]
    \centering
\includegraphics[width=0.5\textwidth]{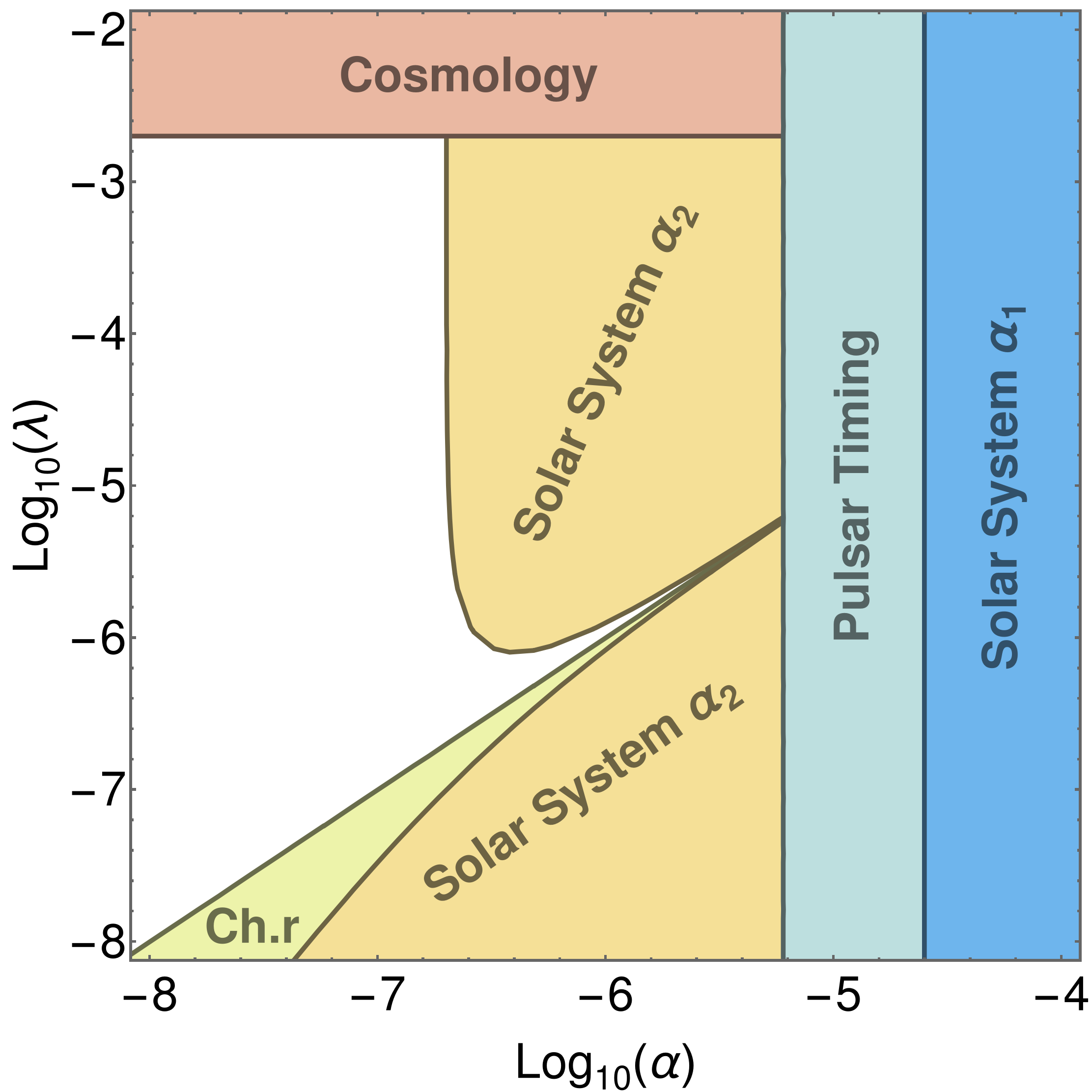} 
\caption{Parameter space of the khrono-metric model in the $(\a,\l)$ plane. We set $\beta=0$ due to the strong constraint (\ref{betabound}). The plot shows the regions excluded by cosmology, Eq.~(\ref{lambdabound}); Solar System tests, Eq.~(\ref{PPNcons}); 
pulsar timing, Eq.~(\ref{alphaPT}); and absence of Cherenkov radiation by cosmic rays, Eq.~(\ref{cchibound}). The couplings $\a$, $\l$ are positive due to the stability bound (\ref{stab}).} 
\label{sensvalues}
\end{figure}

In addition, there are one-sided bounds. The positivity of
the kinetic and gradient energies needed for theoretical consistency
implies 
\be
\label{stab}
\a>0~,~~~~ \b+\l>0\;,
\ee
whereas the absence of Cherenkov radiation of khronon by cosmic rays 
requires \cite{Elliott:2005va}
\be
\label{cchibound}
c_\chi\geq 1\;,
\ee
where $c_\chi$ is the khronon speed (\ref{cchi}).

We summarize the observational bounds on the model parameters in Fig.~\ref{sensvalues}. Since $\b$ is much tighter constrained than the other parameters, see Eq.~(\ref{betabound}), we set it to zero and show the bounds in the $(\a,\l)$ plane. 

\section{Static khronon black hole}
\label{sec:stationary}

In this section we review the structure of static spherically
symmetric black holes in the decoupling limit of the khrono-metric
model following Ref.~\cite{blas_horava_2011}. 
Since in spherical symmetry any vector is hypersurface-orthogonal,
these solutions can also be viewed as spherical black holes in the
Einstein--aether theory \cite{barausse_black_2011}.

Neglecting the khronon EMT, the metric is given by the usual
Schwarzschild solution which we write in 
Finkelstein coordinates to avoid the coordinate singularity at the
matter horizon,
\be
\label{metrFink}
\dd s^2=\left(1-\frac{r_s}{r}\right)\dd v^2-2\dd v\,\dd r -r^2\dd\Omega^2\;,
\ee
where $\dd\Omega^2=\dd \theta^2+\sin^2\theta\dd\psi^2$ is the metric
of a unit sphere and $r_s=2 G_Nm_{\rm bh}$ is the Schwarzschild radius. In
what follows we will also make extensive use of an inverse radial
coordinate,
\be
\label{xidef}
\xi\equiv r_s/r\;,
\ee
leading to the metric
\be
\label{metrxi}
\dd s^2=\left(1-\xi\right)\dd v^2+\frac{2r_s}{\xi^2}\dd v\,\dd \xi 
-\frac{r_s^2}{\xi^2}\dd\Omega^2\;.
\ee

Stationarity of the solution implies that all components of the aether
vector $u_\mu$ are independent of time $v$. We further impose
spherical symmetry, so that the only non-vanishing components are 
\begin{equation}
\label{detail:U_V_notation}
    U \equiv u_v~,~~~~~ V\equiv u^r\;.
\end{equation} 
Note the positioning of the indices in these
definitions. They are convenient since the chosen components
remain invariant under changes of the time coordinate of the form 
$\tilde v =v+f(r)$. In particular, they are the same in the
Finkelstein and Schwarzschild frames, whose time variables are related
by 
\be
\label{vr}
v=t_{\rm Sch}+r+r_s\ln\left(\frac{r_s}{r}-1\right)\;.
\ee 
This choice also simplifies the unit-norm constraint,
\begin{equation}
\label{eqn:unit_norm}
   U^2-V^2=1-\xi\;.
\end{equation}
The fields must satisfy the
boundary conditions,
\be
\label{bcinf_bgr}
U\to 1~,~~~V\to 0~~~~\text{at}~~~~ r\to \infty~~(\xi\to 0)\;,
\ee
expressing that the configuration smoothly merges to the static aether
at infinity. 
Using the reparameterization freedom (\ref{reparam}) the 
khronon configuration can be taken in the form,
\begin{equation}
\label{khronon_bgr}
    \vf = v + r_s \hat\vf(\xi)\;.
\end{equation}
The function $\hat\vf(\xi)$ is
related to the aether components by\footnote{This expression together
  with 
  (\ref{bcinf_bgr}) implies that $\hat\vf(\xi)$ diverges at $\xi\to 
0$ as $-1/\xi$. This is an artifact of the Finkelstein frame whose 
time coordinate $v$ differs from the asymptotic time by a
growing function of the radius.
In the Schwarzschild frame, the condition (\ref{bcinf_bgr}) leads to 
$\vf=t_{\rm Sch}$ at $r\to
\infty$, as it should be. }
\be
\label{fUV}
\hat\vf'(\xi)=\frac{u_\xi}{r_s u_v}=\frac{(U+V)}{\xi^2(1-\xi)U}\;,
\ee
where prime denotes derivative with respect to $\xi$.

The khronon equation (\ref{khrononeqm}) simplifies. First, since the
Schwarzschild metric is Ricci flat, the last term in the khronon
current vanishes. Second, the current conservation equation reduces to 
\begin{equation}
\label{eqn:currents_expanded}
  \partial_r J^r+ \frac{2}{r}J^r=0\;, 
\end{equation}
where we have used stationarity and spherical symmetry. This has a
solution $J^r=C_1/r^2$ with constant $C_1$. 
However, it is straightforward to see that a
non-zero value of $C_1$ is incompatible with the condition
(\ref{bcinf_bgr}) at spatial infinity, implying that the khronon
equation takes the form,
\begin{equation}
\label{Jzero}
    J^r = 0\;.
\end{equation}
Substituting here the expression for the current in terms of the
aether yields
\begin{equation}
\label{khreq_bkg1}
    \frac{U''}{U} - c_\chi^2 \left(\frac{V''}{V} - \frac{2}{\xi^2}\right) = 0\;.
\end{equation}
Finally, using
the relation\footnote{We choose the negative root for $V$ which
  describes the aether tilted towards the center of the black hole.}
 (\ref{eqn:unit_norm}) we obtain a second-order equation
for a single function $U$,
\begin{equation}\label{ude}
	U''+ \frac{c_\chi^2 U}{U^2(1-c_\chi^2) -1 + \xi} \left[
          -(U')^2 + \frac{(UU'+1/2)^2}{U^2-1+\xi} +
          \frac{2(U^2-1+\xi)}{\xi^2} \right]=0\;. 
\end{equation}

The denominator in the second term in (\ref{ude}) 
vanishes at a point $\xi_c$ where
\begin{equation}
\label{Uxic}
    U(\xi_c) = \sqrt\frac{1-\xi_c}{1-c_\chi^2}\;.
\end{equation}
It corresponds to the khronon ``sound'' horizon, i.e. the boundary of
the region from which the khronon waves can escape to infinity. 
It lies outside the metric horizon, $\xi_c<1$, for subluminal khronon,
$c_\chi<1$, and inside the metric horizon, $\xi_c>1$, in the
superluminal case, $c_\chi>1$; when the khronon speed tends to one,
$c_\chi\to 1$,
the metric and khronon horizons coincide, $\xi_c\to 1$. 
We want the solution to be
regular at $\xi=\xi_c$ which implies that the expression in the square
brackets in Eq.~(\ref{ude}) must vanish at this point. This fixes the
first derivative of the aether,
\begin{equation}
\label{Uprimxic}
    U'(\xi_c) = \frac{1}{2(1-c_\chi^2)} \left[
-\sqrt\frac{1-c_\chi^2}{1-\xi_c} 
    +c_\chi \sqrt{\frac{1-c_\chi^2}{1-\xi_c}
-\frac{8c_\chi^2(1-\xi_c)}{\xi_c^2}}\right]\;,
\end{equation}
where we have chosen the root which is
regular at $c_\chi=1$.

\begin{figure}[t]
\begin{center}
\includegraphics[scale=0.6]{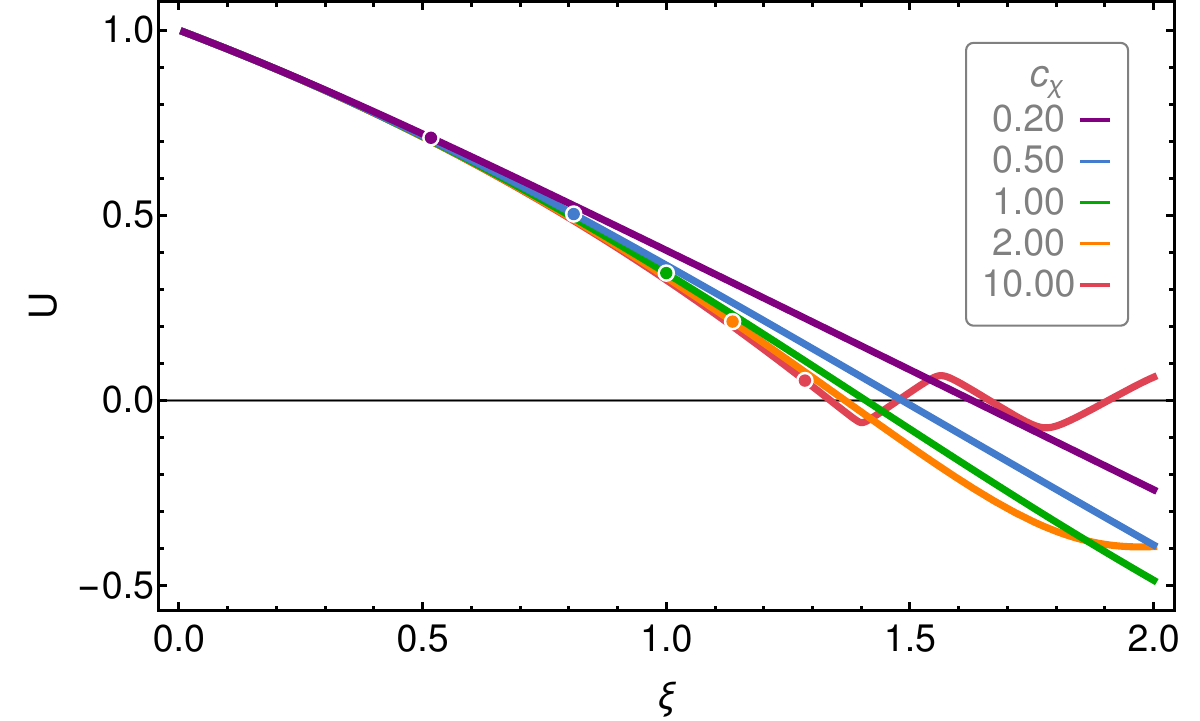}
\end{center}
\caption{The temporal component of the aether vector $U\equiv u_v$
  describing spherically symmetric khronon black hole as a function of
  the inverse radius $\xi\equiv r_s/r$ for several values of the
  khronon speed $c_\chi$. Dots mark the positions of the khronon sound
  horizon $\xi_c$. The first point where a curve crosses zero
  corresponds to the universal horizon $\xi_\star$.
\label{UdiffCx}} 
\end{figure}

We search for the numerical solution of Eq.~(\ref{ude}) by the
shooting method. We choose a trial value $\xi_c$ and determine
the field and its derivative using the boundary conditions
(\ref{Uxic}), (\ref{Uprimxic}). Then we integrate (\ref{ude}) towards
$\xi=0$. Generally, the solution diverges, so we iteratively adjust
$\xi_c$ to get rid of the divergence and satisfy $U(0)=1$. In
practice, both $\xi=\xi_c$ and $\xi=0$ being singular points of the
equation, we set up the boundary conditions at a small distance
$\epsilon\sim 10^{-3}$ away from them, i.e. at $\xi=(\xi_c-\epsilon)$
and $\xi=\epsilon$. The values of the function $U$ and its
derivative at the new points are obtained by
assuming regular Taylor expansions at $\xi=\xi_c$ and $\xi=0$ and
determining their coefficients from Eq.~(\ref{ude}).  
Once a regular solution in the interval
$\epsilon<\xi<(\xi_c-\epsilon)$ is found, we analytically continue it by
Taylor expansion to the other side of the sound horizon and
numerically solve at $\xi>(\xi_c+\epsilon)$. This procedure is repeated
for several values of the khronon speed $c_\chi$ and the resulting
solutions are shown in Fig.~\ref{UdiffCx}. They coincide with the results of paper \cite{blas_horava_2011}.

We observe an important property: all solutions cross zero at some
$\xi_\star>\xi_c$. At this point the aether is orthogonal to the
Killing vector $\d_v$. As a consequence, the foliation leaves that extend to spatial infinity
cannot cross the 2-sphere $\xi=\xi_\star$. Therefore the latter forms a trapped surface even for signals propagating
arbitrarily fast in the preferred frame. Hence the khrono-metric black
holes have non-trivial causal structure, despite violation of Lorentz
invariance. The surface $\xi=\xi_\star$ is called the universal horizon. 

In more detail, writing 
$U(\xi)\approx U_\star'\cdot(\xi-\xi_\star)$ in the vicinity of
$\xi_\star$ and using the relation between the khronon
and aether (\ref{fUV}) 
we find,
\be
\label{fuh}
\hat\vf(\xi)\approx
\frac{1}{U_\star'\xi_\star^2\sqrt{\xi_\star-1}}\ln(\xi_\star-\xi)\;. 
\ee 
We see that $\hat\vf$ logarithmically diverges at $\xi\to \xi_\star$
meaning that the universal horizon lies in the infinite future from
the viewpoint of the asymptotic observer. On the other hand, for an
observer falling into the black hole, the universal horizon is regular
and is reached in a finite proper time. This is clear from
the regularity of all aether components in the Finkelstein frame. We
can also see it directly from the khronon solution by exploiting its
reparameterization invariance. Indeed, the change
\be
\label{khruh}
\tilde \vf=\exp
\left[\frac{U_\star'\xi_\star^2\sqrt{\xi_\star-1}}{r_s}\vf\right]
\ee
makes the khronon field analytic across the universal
horizon. Dynamical formation of the universal horizon in a spherical
collapse has been demonstrated in
\cite{Saravani:2013kva,Bhattacharyya:2015uxt,Bhattacharjee:2018nus,Franchini:2021bpt}.\footnote{Solutions shown in
  Fig.~\ref{UdiffCx} have more zero-crossings at $\xi>\xi_\star$
  suggesting an infinite number of nested universal horizons. A similar
  structure was observed to emerge in the dynamical collapse
  simulations \cite{Bhattacharjee:2018nus}. However, these internal universal
  horizons are causally disconnected from spatial infinity and we do
not consider them in this paper.} 

The solution for the function $U$ can be found analytically in two
limiting cases $c_\chi\to 0$ and $c_\chi\to \infty$. In the first limit
we can neglect the second term in Eq.~(\ref{khreq_bkg1}) which gives
\be
\label{c0bkg}
\frac{U''}{U}=0~~~\Longrightarrow~~~U=1-\frac{\xi}{2}\;,
\ee
where the integration constants have been fixed by the boundary
conditions (\ref{bcinf_bgr}). In this case we get the sound and
universal horizons at $\xi_c=0$ and $\xi_\star=2$.

In the opposite limit $c_\chi\to \infty$ we can drop the
first term in Eq.~(\ref{khreq_bkg1}) and obtain
\be
\label{cinfbkg}
\frac{V''}{V}-\frac{2}{\xi^2}=0~~~\Longrightarrow~~~
V=-b\xi^2~,~~
U=\sqrt{1-\xi+b^2\xi^4}\;.
\ee
The constant $b$ is fixed by requiring existence of a 
universal horizon with $U(\xi_\star)=0$ and
$U'(\xi_\star)$-finite. This gives
\be
\label{bval}
b=\frac{3\sqrt{3}}{16}\;.
\ee
In this case the sound and universal horizons coincide,
$\xi_c=\xi_\star=4/3$.

The characteristics of khronon black holes for various values of the
khronon speed are summarized in Table~\ref{tab:bkg}.

\begin{table}[ht]
\begin{center}
\begin{tabular}{|c|| c | c | c | c | c | c |c |c|c|}
 \hline
$c_\chi$ & 0 &0.2&0.5&0.75&1&1.5&2&10&$\infty$\\[0.5ex]
\hline
$\xi_c$&0&0.52&0.81&0.93&1&1.09&1.14&1.28&1.33\\[0.5ex]
\hline
$\xi_\star$&2&1.63&1.48&1.44&1.41&1.38&1.37&1.34&1.33\\[0.5ex]
\hline
$U'_\star$&-0.5&-0.65&-0.76&-0.82&-0.86&-0.91&-0.93&-1.03&-1.06\\[0.5ex]
\hline
  \end{tabular}
    \caption{Parameters of spherical khronon black holes for several
      values of the khronon speed $c_\chi$: position of the sound
      horizon $\xi_c$, position of the universal horizon $\xi_\star$,
      the slope of the temporal aether component at the universal
      horizon $U'_\star$.
\label{tab:bkg} }
\end{center}
\end{table}

\section{Slowly moving black hole}\label{sec:movingbh}

\subsection{Velocity correction}\label{ssec:movingEOM}

We now turn to our main problem of interest --- a black hole moving
with constant velocity ${\bf v}$ with respect to the preferred
frame. Similar to Ref.~\cite{ramos_constraints_2019}, we assume ${\bf v}$ to be small, ${\rm v}\ll 1$, which
allows us to treat the velocity correction to the khronon field and
the metric at linear order in ${\rm v}$. 

While the presence of khronon breaks the Lorentz symmetry, the action
(\ref{eqn:khrononaction}) is still Lorentz invariant. This means that,
similar to Lorentz invariant theories, we are allowed to perform a
boost on a solution of equations of motion and obtain another
solution. In our case it is convenient to work in the rest frame of
the black hole where the metric and the aether field are static. Of
course, the transformation to the rest frame of the black hole modifies the behavior of khronon at infinity, so the
khronon configuration in the rest frame of the black hole is not
spherically symmetric. The new asymptotic behavior of khronon is
easiest to infer in the Schwarzschild coordinates. Assuming that the
black hole moves along the $z$-axis, the time coordinate before the
boost is related to the Schwarzschild time as
\be
\label{ttSch}
t=\frac{t_{\rm Sch}+{\rm v} z}{\sqrt{1-{\rm v}^2}}\approx 
t_{\rm Sch}+{\rm v} r\cos\theta\;,
\ee
where in the second equality we have kept only the term linear in
${\rm v}$. 
Since $t$ is the time coordinate in the khronon rest frame, we have
$\vf=t$ at $r\to \infty$, which in the rest frame of the black hole
translates into 
\be
\label{vfasymp}
\vf=t_{\rm Sch}+{\rm v}r\cos\theta~~~\text{at}~~r\to\infty\;.
\ee
We see that far from the black hole the khronon acquires a dipole
contribution linear in velocity. Its growth with the radius does not
pose any problems since the corresponding aether field remains
bounded, $u^z=-{\rm v}$ at $r\to\infty$.

The asymptotics (\ref{vfasymp}) suggest the following Ansatz for the
khronon field in the black hole rest frame,
\begin{equation}
\label{eqn:full_khronon}
    \vf = v+r_s\hat\vf(\xi) + {\rm v}r_s \chi(\xi)\cos\theta\;,
\end{equation}
with the function $\hat\vf(\xi)$ from the spherically symmetric
configuration found in Sec.~\ref{sec:stationary}. The function
$\chi(\xi)$ describes the dipolar correction stemming from the
non-zero velocity and obeys the boundary
condition 
\be
\label{chi0bc}
\chi(\xi)\approx \frac{1}{\xi}~~~\text{at}~~\xi\to 0\;.
\ee
The Ansatz (\ref{eqn:full_khronon}) is compatible with the khronon
equations at linear order in ${\rm v}$ since 
at this order the dipolar 
perturbations decouple from perturbations in other multipole sectors
due to the
spherical symmetry of the background solution. The corresponding
linear aether components have the form
\be
\label{aepert}
\delta u_v={\rm v}\,\xi^2 U^2 V\chi'\, \cos\theta~,~~~~
\delta u^r={\rm v}\,\xi^2 U^3 \chi'\, \cos\theta~,~~~~
\delta u_\theta=-{\rm v}r_s\, U\chi\, \sin\theta\;.
\ee
Note that the perturbation 
$\delta u^r=-(r_s/\xi^2)\delta u^\xi$ is finite at spatial
infinity, so we prefer to work with it instead of 
$\delta u^\xi$.

The equation for $\chi(\xi)$ can be obtained by linearizing the khronon
equation (\ref{khrononeqm}). Alternatively, one can find  
the quadratic action for
$\chi$ and then derive the equation from it. We adopt this second path.
Substituting the form (\ref{eqn:full_khronon}) into the
khronon part of the action (\ref{eqn:khrononaction}) and keeping terms
up to quadratic order in $\chi$, after a somewhat lengthy but straightforward
calculation, we obtain  
\begin{equation}\label{perturbationaction}
	S = \frac{2\pi}{3} M^2r_s {\rm v}^2 
\alpha \int \dd v \dd \xi \left[A(\xi)(\chi'')^2 + 
	B(\xi) (\chi')^2 + C(\xi)\chi^2 \right]\;,
\end{equation}
with the coefficients: 
\bseq
\label{ABC}
\begin{align}
	A(\xi) = &U^4\xi^4(V^2 - c_\chi^2U^2)\;,
\label{Aexpr}\\
	B(\xi) = &2\xi^2U^2V^2 - \xi^4U^4VV''
        -\xi^4U^5U'' - 4\xi^4U^3V^2U'' -2\xi^4U^2V^2U'^2 \notag\\ 
		      &- 4\xi^3U^4VV' - 8\xi^3U^3V^2U' -
                      4\xi^4U^3VU'V' - 2\xi^2U^4V^2 
	\notag\\
	&- c_\chi^2 ( 4\xi^2U^4 - 3\xi^4U^4VV''
        -3\xi^4U^5U''-6\xi^4U^4U'^2 - 12\xi^3U^5U' + 6\xi^2U^4V^2)\;,
\label{Bexpr}\\ 
	C(\xi) = &2\big[ -\xi^2U^3U'' -\xi^2UV^2U''
        -\xi^2U^2U'^2 -2\xi UV^2U' -2\xi^2UVU'V'\notag\\ 
	&+ c_\chi^2 (\xi^2U^2VV'' + \xi^2U^3U'' +
                      3\xi^2U^2U'^2 + 10\xi U^3U' -2\xi U^2) \big] \;.
\label{Cexpr}
\end{align}
\eseq
In this derivation we have used the relations (\ref{eqn:unit_norm}), (\ref{fUV})
and performed integrations
by parts to simplify the final result.
The expression (\ref{perturbationaction}) coincides, up to an overall
normalization factor, with the special case $\ell=1$ of the 
action obtained in
\cite{blas_horava_2011} for perturbations in the 
$\ell$th multipole sector.\footnote{Ref.~\cite{blas_horava_2011} considers
  time-dependent perturbations and works in the frame where the time
  coordinate is identified with the background khronon field because
  only in this frame the evolution of perturbations is
  self-consistent. We do not need to perform this change of variables
  since we focus on static perturbations.}  
We derive from it the equation
\begin{equation}\label{perturbationEOM}
    (A \chi'')'' - (B \chi')' + C\chi = 0.
\end{equation}
Equation (\ref{perturbationEOM}) is a fourth order differential equation, consistent with the fact
that khronon action contains higher derivatives.
Our task is to find a solution of this equation satisfying the
boundary condition (\ref{chi0bc}) and regular at $\xi\leq\xi_\star$.

\subsection{Regularity conditions}
\label{sec:moving}

The equation (\ref{perturbationEOM}) has three singular points where
the coefficient $A(\xi)$ in front of the highest-derivative term
vanishes. They correspond to the spatial infinity at
$\xi=0$, the khronon sound horizon $\xi=\xi_c$, and the universal
horizon $\xi=\xi_\star$. We now derive the conditions on the
solution following from the 
requirement that the khronon foliation is
regular at these points. Note that the singular points do not
include the Schwarzschild horizon $\xi=1$ since we do not consider the
equations for the metric perturbations at the leading order in the
decoupling limit.

According to the general rules, we Taylor expand the coefficients $A$,
$B$, $C$ in the vicinity of the singular points and look for the
solution in the form of a power-law $\chi\propto (\xi-\xi_0)^\g$,
where $\xi_0=0,\;\xi_c$ or $\xi_\star$. For $\xi_0=0$ we obtain the
following powers of the independent solutions:
\begin{equation}
\label{powersxi0}
	\g_1^{(0)}=-3\;,~~~ \g_2^{(0)} = -1\,,~~~  
\g_3^{(0)} = 0\; ,~~~ \g_4^{(0)}=2\;.
\end{equation}
On the other hand, the boundary condition at infinity implies the
behavior (\ref{chi0bc}). We see that the solution corresponding to
$\g_1^{(0)}$ must be rejected.

At the point $\xi_0=\xi_c$ the powers are: 
\begin{equation}
\label{powersxic}
	\gamma_1^{(c)} = 0\;,~~~
\gamma_2^{(c)}=\g_3^{(c)} = 1\;,~~~\gamma_4^{(c)} = 2\;. 
\end{equation}
The fact that two powers coincide implies that the power-law Ansatz
yields only three linearly independent solutions. The fourth solution
has a logarithmic factor and behaves as
\be
\label{logxic}
(\xi-\xi_c)\log(\xi-\xi_c)\;.
\ee
This is not analytic at $\xi=\xi_c$. Moreover, it leads to a
divergence in the components of the 
linear aether perturbation (\ref{aepert}) which contain the radial
khronon derivative.
Thus, for the foliation to be regular at the khronon sound horizon, we
have to reject the solution (\ref{logxic}).

Next, at the universal horizon, $\xi_0=\xi_\star$, we have four
independent power-law solutions with 
\begin{equation}
\label{powerUH}
 \g_{1}^{(\star)} = -\frac{1}{2} - \sqrt{\frac{1}{4}+\frac{2}{\xi_\star^2
     {U'_\star}^2}}\;,~~~
  \g_2^{(\star)} = -1\;,~~~ 
\g_3^{(\star)} = 0\;,~~~
\g_4^{(\star)} = 
-\frac{1}{2} + \sqrt{\frac{1}{4}+\frac{2}{\xi_\star^2 {U'_\star}^2}}\;.
\end{equation}
Again, we require the aether perturbations (\ref{aepert}) to be finite
which implies that $\g$ must be larger or equal $-1$. This rejects the
solution corresponding to $\g_1^{(\star)}$, while the other solutions
are admissible. Note that $\g_4^{(\star)}$ is non-integer and
according to Table~\ref{tab:bkg} lies in the interval
\be
\label{g4range}
0.62\leq \g_4^{(\star)}\leq 1\;.
\ee
The corresponding perturbations of the aether (\ref{aepert})
vanish at the universal horizon, together with their first
derivatives. The second derivatives of
$\delta u_\mu$, however, diverge, so the
solution has non-analyticity at $\xi=\xi_\star$. Nevertheless, we show
in Appendix~\ref{stressenergy} that the divergent terms cancel in the
linearized khronon EMT and it remains finite at the universal
horizon. We will return to the discussion of the non-analytic khronon
behavior in Sec.~\ref{sec:conc}. 

To summarize, we have found that at each singular point of
Eq.~(\ref{perturbationEOM}) there are four linearly independent
solutions, one of which must be rejected to ensure the regularity of
the foliation. This imposes three conditions on the function
$\chi$. Supplementing them with the normalization of the $\xi^{-1}$
term at spatial infinity (\ref{chi0bc}) we obtain four conditions
which match the number of free parameters in the general solution of a
fourth order differential equation. Since the equation is linear and
barring accidental degeneracies, one expects existence of a unique
solution 
satisfying the regularity conditions in the whole interval
$0<\xi<\xi_\star$ for every value of the khronon speed $c_\chi$.
In
the next subsection we confirm this expectation by explicitly
constructing such solutions numerically.

\subsection{Numerical solutions}
\label{sec:solutions}

\begin{figure}[t]
    \centering
\includegraphics[scale=0.6]{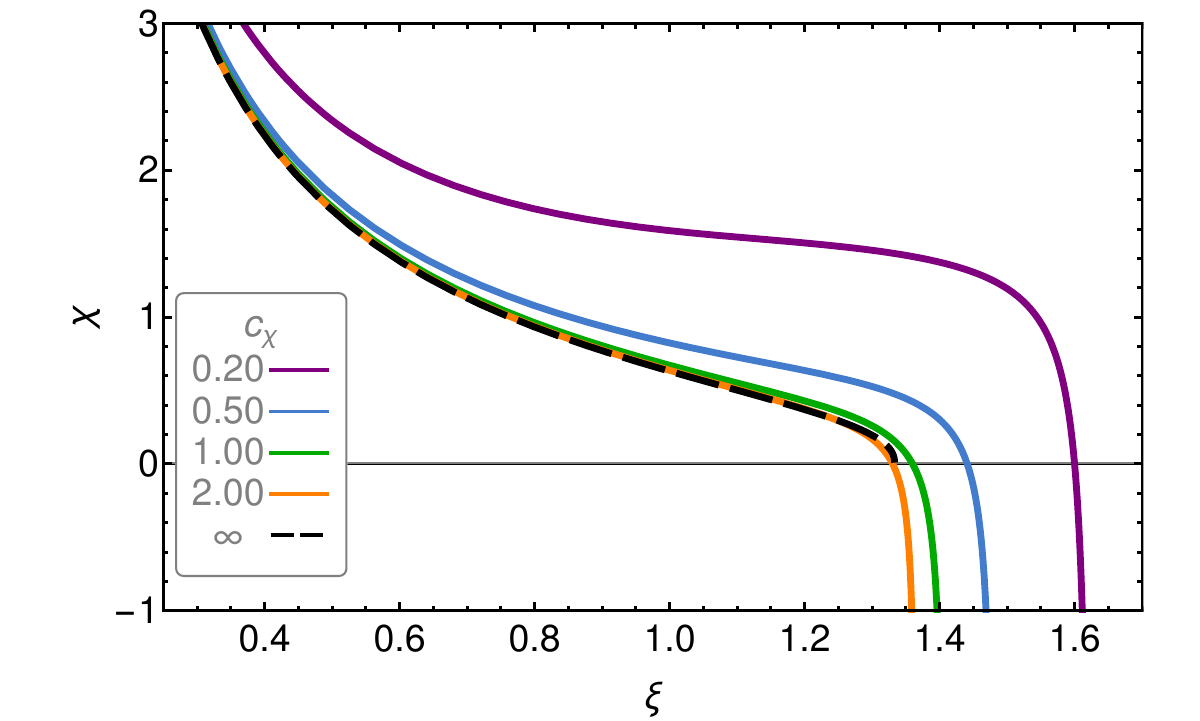}
\caption{Khronon perturbation in the rest frame of a slowly moving black hole
for different values of $c_\chi$ as a function of the inverse radial
coordinate $\xi\equiv r_s/r$.} 
\label{ChidiffCx}
\end{figure}

\begin{figure}[t]
    \centering
\includegraphics[scale=0.55]{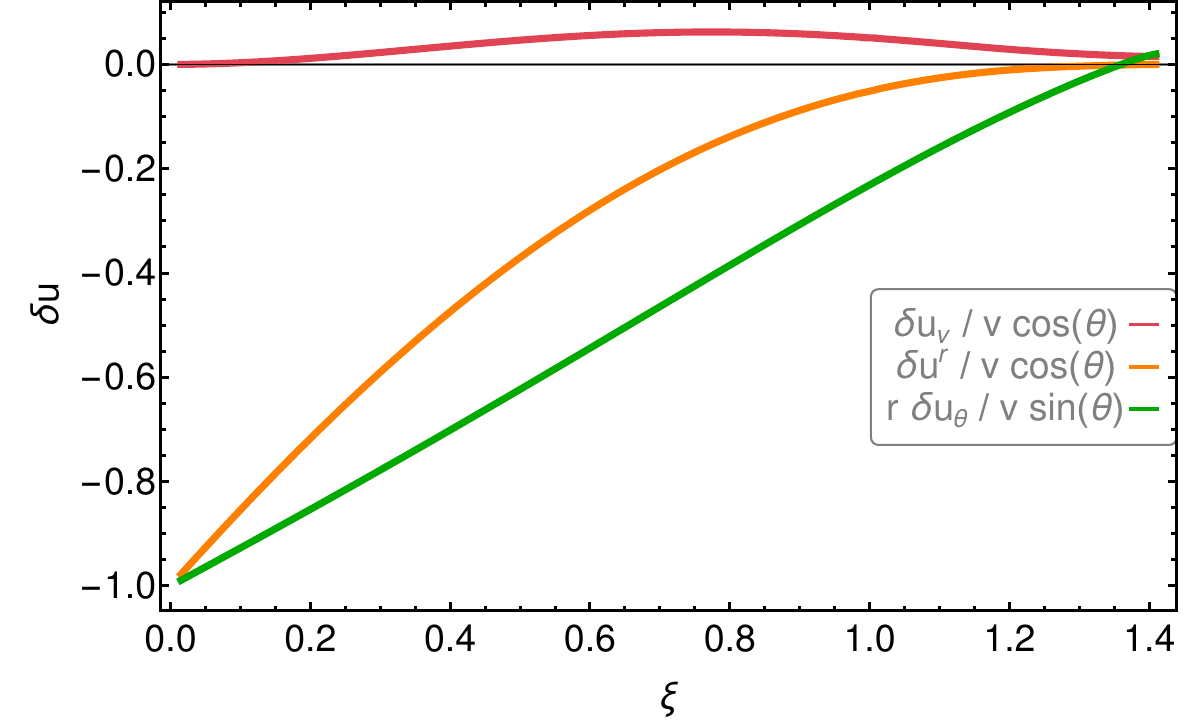}
\caption{The dependence of the perturbed 
aether components on the inverse
  radius $\xi\equiv r_s/r$. 
The angular component $\delta u_\theta$ is multiplied by $r$ to have a
finite value at spatial infinity. 
The khronon speed is taken to be
$c_\chi=1$.}
\label{fig:ws}
\end{figure}

To find the global solution of Eq.~(\ref{perturbationEOM}) in the
range $0<\xi<\xi_\star$ we use the
following algorithm. At each of the three points
$\xi_0=0,\;\xi_c,\,\xi_\star$ we construct a linear combination of the
three admissible solutions. The coefficient of the $1/\xi$ solution at
$\xi_0=0$ is further fixed to unity according to the normalization
(\ref{chi0bc}). The $8$ remaining coefficients are treated as unknowns. We
numerically integrate the equation away from the singular points and
match the solutions through their third derivatives
in the middle of the intervals $(0,\xi_c)$,
$(\xi_c,\xi_\star)$. This
provides $8$ linear equations which are solved to determine the $8$
unknown coefficients. The details of the algorithm are explained in
Appendix~\ref{combiningChi}. The resulting solutions are shown in 
Fig.~\ref{ChiTildediffCx} for several values of the khronon speed
$c_\chi$. We see that they are smooth everywhere inside the interval
$0<\xi<\xi_\star$.  

At $\xi=0$ and $\xi=\xi_\star$ the solutions behave as $\xi^{-1}$ and
$(\xi-\xi_\star)^{-1}$, as expected from the analysis of
Sec.~\ref{sec:moving}.\footnote{Except for the curve with $c_\chi=\infty$
  which reaches zero at the universal horizon,
  see below.} As explained there, this behavior corresponds to a
regular foliation. To show this explicitly, we plot the linear
perturbations of the aether components (see Eqs.~(\ref{aepert})) in
Fig.~\ref{fig:ws} for a representative value of $c_\chi$. 
We observe that they are finite both in the asymptotic region 
($\xi=0$) and at the universal horizon. We have also checked that the
components of the khronon EMT are finite on the solution.

\subsection{Limit $c_\chi= \infty$}
\label{sec:infcchi}

It is instructive to consider the limit of infinite khronon speed. As
we are going to see, the equation for the velocity-induced correction
$\chi$ greatly simplifies in this case. Solving it with an independent
method allows us to cross-check our general results.

Let us go back to Eq.~(\ref{khrononeqm}). The limit $c_\chi=\infty$
is achieved by setting $\a=0$, so that the first term in the
khronon current disappears. Recalling also the Ricci-flatness of the
Schwarzschild metric, we see that the equation takes the form, 
\begin{equation}
\label{cinfueq}
\tilde\Delta \vartheta = 0\;,
\end{equation}
where $\tilde\Delta$ is an elliptic operator,
\be
\label{tildeDel}
\tilde \Delta\equiv \nabla_\nu\, X^{-1/2} P^\nu_\mu \nabla^\mu\;.
\ee
In flat spacetime with khronon proportional to the time coordinate
this operator becomes a spatial
Laplacian. In general, it represents a curved-space version of the
Laplacian acting on the leaves of the preferred foliation. Similar to
the standard Laplacian, one can show that it has only
constant zero modes (see Appendix~\ref{app:Lapl}). 
Then Eq.~(\ref{cinfueq}) implies $\vartheta={\rm const}$ and recalling
that $\vartheta=\nabla_\mu u^\mu$ must vanish at spatial infinity we
arrive at a much simpler khronon equation,
\begin{equation}
\label{cxlimit}
    \nabla_\mu u^\mu=0\;.
\end{equation}
It is straightforward to check that the background khronon
configuration (\ref{cinfbkg}) satisfies this equation. Considering the
linear dipolar perturbation (\ref{eqn:full_khronon}), we obtain
\be
\label{cinfchieq}
\left(1-\xi+\frac{27}{256}\xi^4\right)\chi''-
\left(\frac{3}{2}-\frac{81}{128}\xi^3\right)\chi'-\frac{2}{\xi^2}\chi=0\;,
\ee 
where we have used the background expressions (\ref{cinfbkg}),
(\ref{bval}). Thus, we have reduced the order of the differential
equation for $\chi$ from 4 to 2.

\begin{figure}[t]
    \centering
\includegraphics[scale=0.55]{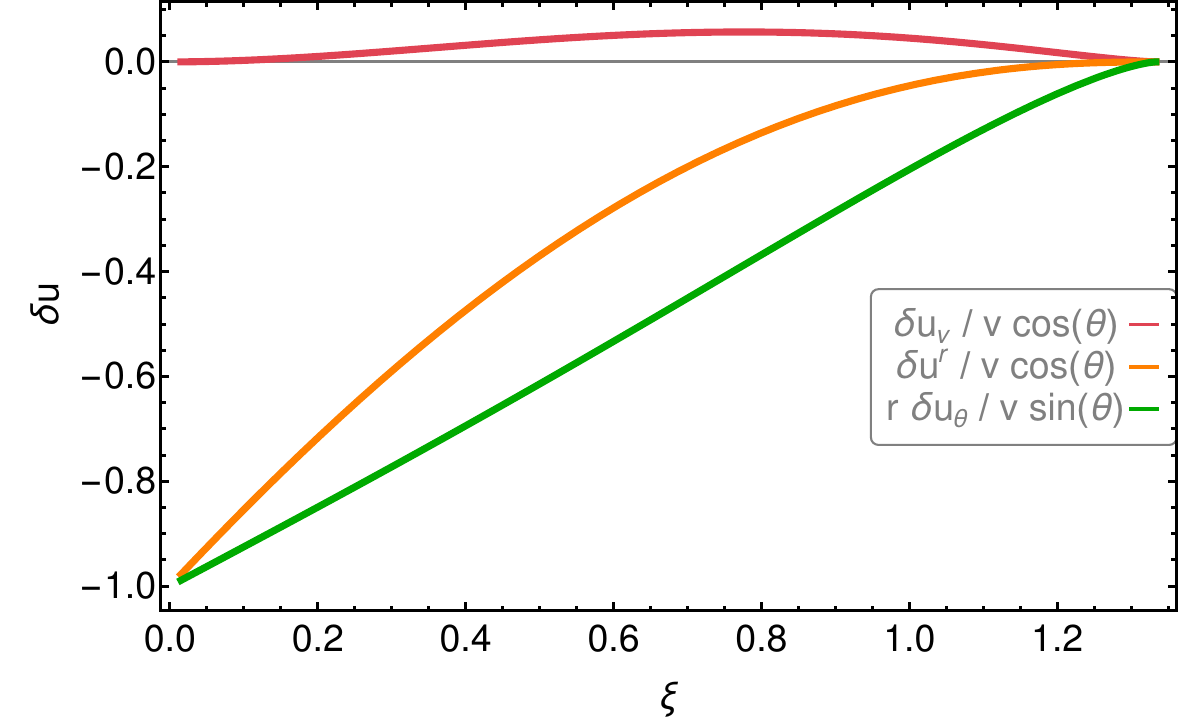}
\caption{Perturbed 
aether components for the case of infinite khronon speed, $c_\chi=\infty$. 
 The configuration is very similar to that 
 for $c_\chi=1$, with notable difference only in the vicinity of the universal horizon $\xi_\star=4/3$,
 cf. Fig.~\ref{fig:ws}. }
\label{fig:cinfu}
\end{figure}

The singular points of Eq.~(\ref{cinfchieq}) lie at $\xi=0$ and
$\xi=4/3$. The latter corresponds to the coincident universal and
sound horizons. Using the
power-law Ansatz $\chi\propto (\xi-\xi_0)^\gamma$ at these points, we
obtain the following powers:
\bseq
\label{cinfpowers}
\begin{align}
&\xi_0=0\;:&&\gamma^{(0)}_1=-1~,&&\gamma_2^{(0)}=2\;,\\
&\xi_0=4/3\;:&&\gamma^{(\star)}_1=-1-\sqrt{2}~,&&
\gamma_2^{(\star)}=-1+\sqrt{2}\;. 
\end{align}
\eseq
The solution with $\gamma_1^{(\star)}$ must be rejected since it
corresponds to the aether divergent at the universal horizon. All
other solutions are admissible. 
This suggests to start the numerical integration of
Eq.~(\ref{cinfchieq}) from the vicinity of the 
point $\xi_0=4/3$ where the solution is
uniquely fixed, modulo an overall normalization. Once the solution in
the range $0<\xi<4/3$ is obtained, we normalize it by the condition
(\ref{chi0bc}). The result is shown in Fig.~\ref{ChidiffCx} by the black dashed
curve. We have checked that it represents a limit of the solutions with
finite $c_\chi$ when this parameter increases. The limit in the
vicinity of the universal horizon is not uniform, since all solutions
with $c_\chi<\infty$ diverge there, whereas the 
$c_\chi=\infty$ solution vanishes (though in a non-differentiable
way). 

For completeness, we also show in Fig.~\ref{fig:cinfu} the components
of the aether vector corresponding to the $c_\chi=\infty$ solution. They are very similar to the aether configurations with finite $c_\chi\geq 1$, differing slightly only in the vicinity of the universal horizon. 
As
expected, the aether components are finite and continuously differentiable down to the
universal horizon, exhibiting non-analyticity only in the second
derivative. 
They agree with the solutions of Ref.~\cite{ramos_constraints_2019}
obtained in the case $\a=\b=0$.

\section{Sensitivities}
\label{sec:sens}

The breaking of Lorentz invariance leads to violation of the strong
equivalence principle. One manifestation of this violation is the
dependence of the black hole inertial mass on its velocity which is captured by
the sensitivity parameters \cite{foster_strong_2007}.
Non-zero sensitivities affect post-Newtonian dynamics of binary
systems and, in the case of different sensitivities for the two
binary components, can lead to dipolar gravitational radiation
\cite{foster_strong_2007,yagi_constraints_2014}. In this section we
use the solutions obtained above to extract the leading-order
sensitivity of a black hole in the
khrono-metric model. 

At length scales much larger than the Schwarzschild radius a black
hole can be modeled as a point particle with an effective action,
\be
\label{Spp}
S_{\rm pp}=-m\int d\tau\, F(\upgamma)\;,
\ee 
where $\tau$ is the particle's 
proper time and $F$ is a function of the boost factor
$\upgamma={\rm v}^\mu u_\mu$ of the particle with respect to the preferred
frame. Here ${\rm v}^\mu$ is the particle's four-velocity and we
normalize $F(1)=1$, so that $m$ represents the mass of particle at
rest. This is the most general action containing ${\rm v}^\mu$ and
$u^\mu$ without derivatives and compatible with the symmetries of the
theory. A system of several black holes will include several copies of
this action which, combined with the khrono-metric action
(\ref{eqn:action}), will fully describe the long-distance dynamics of
the system. 

In the non-relativistic limit $\upgamma$ is close to 1, so we can
expand
\be
\label{Fexpand}
F(\upgamma)=1+\sigma\,(1-\upgamma)+\frac{\sigma'}{2}(1-\upgamma)^2+\ldots\;,
\ee
where the coefficients $\sigma$, $\sigma'$ are called sensitivities. 
In what follows we focus on the first sensitivity $\sigma$ which
multiplies the correction proportional to the square of the particle's
velocity. Note that so far we have worked at linear order in velocity, so 
it may seem that $\sigma$
is beyond our reach. This problem was overcome in
\cite{foster_strong_2007,yagi_constraints_2014,ramos_constraints_2019}
by noticing that the particle-aether interaction (\ref{Spp}) induces
off-diagonal corrections to the metric $g_{0i}$ which are linear in
${\bf v}$ and can be matched to the asymptotic behavior of the full
non-linear metric. 

While this approach allows one to work at order
$O({\rm v})$, it still requires solving for the khronon-induced
corrections to the metric which presents a challenging task. Here we
use a method that bypasses it. We observe that even in a
fixed metric the interaction (\ref{Spp}) provides a source for the
khronon whose perturbation will therefore depend on
$\sigma$. Thus, the sensitivity can be read off by matching this
perturbation to the asymptotics of the khronon field embedded in the
Schwarzschild background. 

As before, we work in the rest frame of the particle. Neglecting
khronon back-reaction on the metric, the latter can be taken in the
Newtonian gauge,
\be
\label{Newtmetr}
\dd s^2=(1+2\phi)\,\dd t^2-(1-2\phi)\dd {\bf x}^2\;,
\ee
where
\be
\label{Newtpot}
\phi=-\frac{G_N m}{|x|}
\ee
is the Newtonian potential. The khronon is taken in the form
\begin{equation}
\label{sens:view2}
    \varphi = t + {\bf vx} + \delta\vf({\bf x})\;.
\end{equation}
Expanding the point particle action (\ref{Spp}) to linear order in
velocity and khronon perturbation yields the source term,
\be
\label{khsource}
S^{(\delta\vf)}_{\rm pp}=m\sigma\int \dd t\,{\rm v}^i\d_i\delta\vf\;.
\ee
Next, we expand the khrono-metric action (\ref{eqn:action}) and keep
terms linear or quadratic in $\delta\vf$ and up to $O({\rm v})$ in
velocity: 
\be
\label{khspace}
S^{(\delta\vf)}=-\frac{M^2}{2}\int \dd^4
x\Big[(\b+\l)(\Delta\delta\vf)^2
+2(-\a+3\b+\l)\Delta\delta\vf\,{\rm v}^i\d_i\phi\Big]\;,
\ee
where $\Delta=\d_i\d_i$ is the Laplacian. Variation with respect to
$\delta\vf$ gives us the khronon perturbation equation,
\begin{equation}
\label{delatphieq}
   M^2(\b+\l)\Delta^2 \delta\vf +
    M^2(-\alpha+3\beta+\l)\,{\rm v}^i \partial_i \Delta\phi +
  m\sigma\,{\rm v}^i\prd_i\delta({\bf x})=0\;,
\end{equation}
which is straightforward to solve using the Fourier
decomposition. Combining the answer into the full khronon
configuration we obtain,
\be
\label{khronfin}
\vf=t+{\rm v}|x|\cos\theta +{\rm v}
G_Nm\frac{(-\a+3\b+\l+2\sigma)}{2(\b+\l)}\cos\theta\;, 
\ee
where we have used the relation (\ref{GNewt}) and chosen the $z$-axis
along the direction of velocity.

We want to match this result to the khronon solution found in
Sec.~\ref{sec:movingbh}. This requires some care since 
the Schwarzschild
radial coordinate $r$ used there differs from $|x|$. The easiest way
to see this is by comparing the angular parts of the metric:
\be
\label{rxrel}
r^2\dd \Omega^2=(1+2\phi)|x|^2\dd \Omega^2
~~~\Longrightarrow~~~|x|=(1-\phi)r=r-G_Nm\;.
\ee
Substituting into (\ref{khronfin}) and using $G_Nm=r_s/2$ we finally
obtain, 
\be
\label{khronfin1}
\vf=t+{\rm
  v}r_s\bigg(\frac{r}{r_s}+\frac{(-\a+\b-\l+2\sigma)}{4(\b+\l)} 
\bigg)\cos\theta\;.
\ee
The expression in brackets matches to the first two terms in the
expansion of the function $\chi(\xi)$ introduced in
(\ref{eqn:full_khronon}),\footnote{Recall that $\xi=r_s/r$.} 
\begin{equation}
\label{chilead}
    \chi =  \frac{1}{\xi} + \chi_0 +O(\xi)\;.
\end{equation}
This allows us to connect the sensitivity to the subleading
coefficient in the asymptotic behavior of this function,
\begin{equation}
\label{sigmaexp1}
    \sigma = \frac{\alpha -\b+ \lambda}{2}+ 2 (\beta+\lambda) \chi_0(c_\chi)\;,
\end{equation}
where we have explicitly indicated the dependence of $\chi_0$ on the
khronon speed (\ref{cchi}). 
This result agrees with the $\a,\b,\l\ll 1$ limit of the equation in 
\cite{ramos_constraints_2019}
expressing the sensitivity in terms of the asymptotics of the khronon field.\footnote{Note that our coefficient $\chi_0$ is denoted as $\chi_0(G_N\tilde m\delta_0)^{-1}$ in \cite{ramos_constraints_2019}.} However, Ref.~\cite{ramos_constraints_2019} was unable to study the dependence of $\chi_0$ on the khronon couplings since it did not find regular black hole solutions for $\a,\b\neq 0$.   

\begin{table}[t]
        \centering
        \begin{tabular}{|c|| c | c | c | c | c | c |c |c|c|}
           \hline
        $c_\chi$ & 0.2&0.5&0.75&1&1.5&2.0&5.0&10&$\infty$ \\[0.5ex]
        \hline
$\chi_0$&0.2910&-0.1511&-0.2057&-0.2251&-0.2390&-0.2438&-0.2490&-0.2498&-0.2500\\[0.5ex] 
\hline
\end{tabular}
    \caption{Subleading coefficient $\chi_0$ in the asymptotic
      expansion of the khronon perturbation for several values of the
      khronon speed.}
        \label{sequences}\centering
\end{table}

\begin{figure}[t]
    \centering
\includegraphics[scale=0.6]{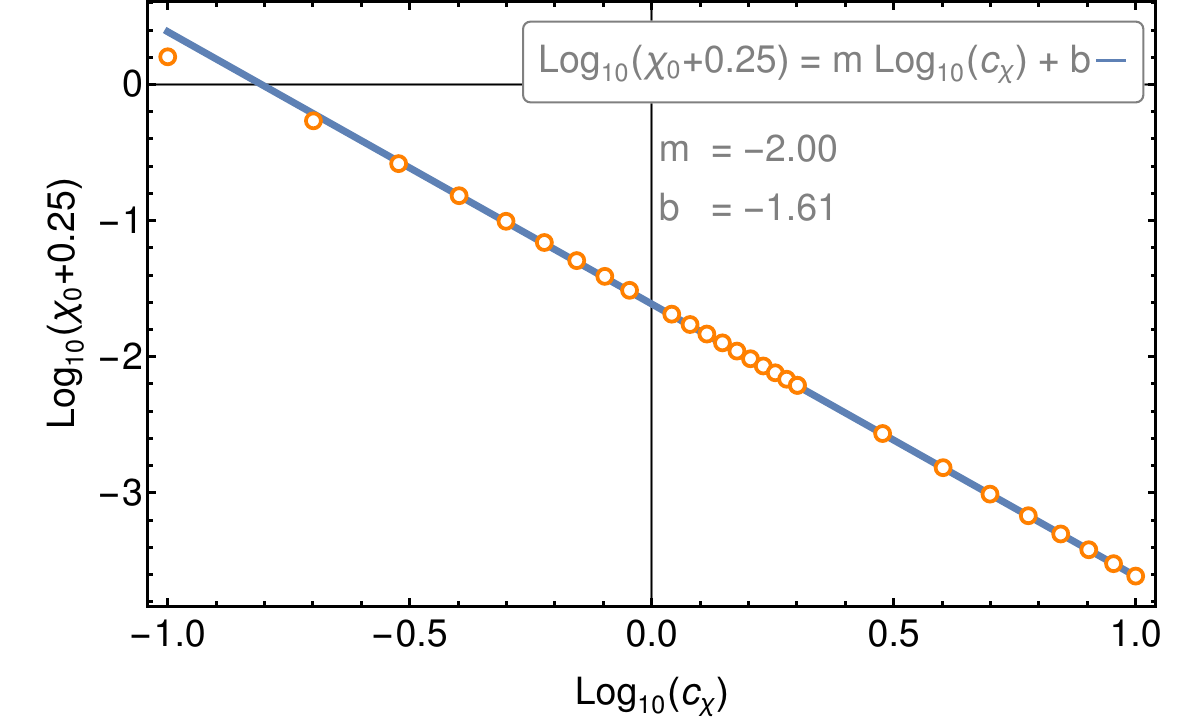}
\caption{Dependence of $(\chi_0+1/4)$ on the khronon speed. Dots
  represent the numerical values, whereas line shows the analytic
  approximation (\ref{kappa}).} 
\label{sensplot}
\end{figure}

The values of $\chi_0$ for several khronon speeds are listed in
Table~\ref{sequences}. 
At large $c_\chi$ we have a limit
\be
\label{chi0lim}
\lim_{c_\chi\to \infty}\chi_0=-1/4\;.
\ee
This follows from substituting the form (\ref{chilead}) into the
khronon equation (\ref{cinfchieq}) valid at $c_\chi=\infty$ and
analyzing its Taylor series at $\xi=0$. At general $c_\chi$ values
$\chi_0$ cannot be determined by a local analysis at $\xi=0$ since it
represents a free parameter of the general solution at this point
(cf. Sec.~\ref{sec:moving}). It is possible to derive correction to
$\chi_0$ in $1/c_\chi$ expansion which has the form (see
Appendix~\ref{app:chi0corr}), 
\begin{equation}
\label{kappa}
    \chi_0 = -\frac{1}{4} + \frac{\kappa}{c_\chi^2}~,~~~~~
\kappa=0.0244\;.
\end{equation}
This expression happens to fit the exact numerical values of $\chi_0$
remarkably well down to $c_\chi\sim 0.1$, as
illustrated in Fig.~\ref{sensplot}. Substituting (\ref{kappa}) into
Eq.~(\ref{sigmaexp1}) and using the expression  (\ref{cchi}) for the khronon speed we obtain, 
\begin{equation}
\label{sigmaexp2}
    \sigma = \frac{\alpha-2\beta}{2} + 2\kappa\, \alpha\;.
\end{equation}
This gives a simple formula for the black hole sensitivity directly in
terms of the model parameters valid at $c_\chi\gtrsim 0.1$.

In the phenomenologically relevant case $|\b|\ll\a$, we have from (\ref{sigmaexp2}) $\s\approx 0.55\a$. Given the existing constraints discussed in Sec.~\ref{ssec:constraints}, the sensitivity happens to be small,
\be
\label{sigmaless}
\s\lesssim 3\cdot 10^{-6}\;.
\ee
Thus detecting the effects proportional to $\s$ will be challenging. It will be interesting to analyze if these effects can be discerned with the exquisite precision of future LISA measurements~\cite{Berry:2019wgg}.

We observe that the sensitivity is essentially independent of the coupling $\l$ as long as $c_\chi\geq 1$. This is to be expected
since in the limit $\a,\b\to0$ the khronon
becomes a ``stealth'' field and does not cause any back-reaction on
the metric \cite{Berglund:2012bu,ramos_constraints_2019,Franchini:2021bpt}.
We can see it directly from the khronon EMT (\ref{khEMT}). Setting
$\a=\b=0$, we obtain that it is proportional to the aether divergence
$\vartheta$. The latter, however, vanishes in the limit
$c_\chi=\infty$, see Eq.~(\ref{cxlimit}). 
This argument applies not only to black holes, but to any asymptotically flat spacetime. 

Finally, the sensitivity is independent of the black hole mass. This means that
Lorentz violation does not lead to dipolar gravitational radiation in
a binary system composed of a pair of Schwarzschild black holes. In this connection, it will be interesting to study if the situation will change for rotating black holes, in which case the sensitivity can depend on the black hole spin. This question is, however, outside the scope of the present work.  
A dipolar
radiation can also arise in black hole -- neutron star or black hole --
white dwarf binaries, since the sensitivities of the two companions
will be in general different. Due to the argument given above, the radiation is expected to be suppressed by the small couplings $\a$ and $\b$. Thus, observations of binary systems are unlikely to improve the bounds on the less constrained coupling $\l$.

Before closing this section, it is instructive to go one step further and consider the
back-reaction of the khronon on the long-distance black hole
metric. Ref.~\cite{yagi_constraints_2014} found that the metric of a
body with sensitivity $\sigma$ contains dipolar and quadrupolar
contributions depending on the body's velocity with respect to the
preferred frame. 
In the decoupling limit the coefficients of these terms read:
\be
\label{hatPPNgen}
\tilde\a_1=-4(\a-2\b-2\sigma)~,~~~~~~~~
\tilde\a_2=\frac{(\a-2\b-2\sigma)(\a-3\b-\l)}{2(\b+\l)}\;.
\ee  
Substituting here the expression (\ref{sigmaexp2}), we obtain
simplification for the black hole case,
\be
\label{hatPPNbh}
\tilde\a_1=16\kappa\,\a~,~~~~~~~~
\tilde\a_2=-\frac{2\kappa\,\a\,(\a-3\b-\l)}{\b+\l}\;.
\ee  
These corrections to the moving black hole metric are analogous to the
dipolar and quadrupolar terms in the post-Newtonian expansion of a
weakly gravitating body entering with the preferred-frame parameters
$\alpha_1^{\rm PPN}$, $\alpha_2^{\rm PPN}$ given by
(\ref{PPNformulas}).\footnote{The preferred-frame
  parameters $\hat\a_1$, $\hat\a_2$ introduced in \cite{Will:2018ont}
  for strongly gravitating binary
  systems reduce to $\tilde\a_1$, $\tilde\a_2$ when one of the
  companions is a test particle with vanishing sensitivity.} 
Setting for concreteness the most constrained coupling $\beta$ to zero, we make two observations. First, the black hole
parameters (\ref{hatPPNbh}) have opposite signs to their weak-field
counterparts. Second, their absolute value is suppressed by about an order
of magnitude due to the smallness of the numerical 
coefficient $\kappa$, see Eq.~(\ref{kappa}). Remarkably,
the gravitational field of a black hole appears to be closer to the GR
prediction than that of a weakly bound object, such as a star or a
planet.

\section{Discussion}
\label{sec:conc}

We have found solutions of the khrono-metric model describing slowly
moving Schwarzschild black holes. We worked at linear order in the
black hole 
velocity with respect to the preferred frame and in the decoupling
limit, i.e. neglecting the effect of the khronon EMT on the
metric. 
The use of the decoupling limit allowed us to reduce the complicated system of equations for the metric and khronon to a single khronon equation in the fixed Schwarzschild spacetime. Expansion in the black hole velocity, in turn, split the task into two steps: solving a second-order ordinary differential equation for the spherically symmetric background, and solving a fourth-order linear ordinary differential equation for the dipole correction proportional to the velocity. We showed that the boundary conditions at infinity, at the khronon ``sound" horizon and at the universal horizon uniquely fix the solutions of both equations and we constructed these solutions numerically. 

The solutions we found are regular everywhere down to the
universal horizon radius~$r_\star$. In particular, they are smooth at
the Schwarzschild and khronon ``sound" horizons lying outside
$r_\star$. However, the solution cannot be unambiguously continued
inside $r_\star$ because of a weak branch point
singularity. Appearance of such singularity is consistent with the
general argument \cite{blas_horava_2011} 
that aspherical perturbations of Schwarzschild
background destroy the analyticity of the universal horizon. This
raises a question about the structure of black holes at
$r<r_\star$ which likely cannot be resolved within the low-energy
khrono-metric model and requires invoking an ultraviolet completion,
such as Ho\v rava gravity. On the other hand, the singularity at the
universal horizon appears sufficiently benign not to destroy 
predictivity in the outer region. Indeed, the universal horizon lies
in the future of all preferred foliation leaves extending to infinity
and the physics on them remains agnostic about the singularity~\cite{Bhattacharyya:2015gwa}. 

Our results differ from those of Ref.~\cite{ramos_constraints_2019} which was unable to find any moving black hole solutions regular down to the universal horizon for non-zero values of the khronon couplings $\a$, $\b$. We do not know the exact origin of the problem but suspect that it is related to the complexity of the full system of metric and khronon equations which was analyzed numerically in \cite{ramos_constraints_2019}. Thus, it would be interesting to redo the analysis of \cite{ramos_constraints_2019} with the insights obtained in our work. In particular, one can start by including the khronon EMT perturbatively, as the source for the deviations of the metric away from the Schwarzschild form. Since khronon is regular at $r>r_\star$, this procedure will not lead to any singularities of the metric outside the universal horizon. Whether the universal horizon itself remains regular is less clear: In principle, the non-analyticity of khronon at $r_\star$ might give rise to a 
curvature singularity. We do not expect this to happen, however, at least in the linear order in $\a$ and $\beta$ since, as we showed, the khronon EMT is bounded at $r_\star$. Further evidence that the universal horizon can remain regular with the full back-reaction of khronon included comes from the recent study \cite{Saito:2024zqh} 
of black holes in a tensor-scalar theory generalizing 
the khrono-metric model with $\a=0$ and
non-zero $\b$ and $\l$. 

Another interesting question concerns implications of our results for black
hole thermodynamics. The temperature of Hawking radiation in Lorentz
violating theories in general depends on the type of emitted particles
which leads to an apparent contradiction with the second law
\cite{Dubovsky:2006vk}. The studies of this issue in Ho\v rava gravity
have been restricted so far to spherical Schwarzschild black holes and
led to controversial results
\cite{Berglund:2012fk,Michel:2015rsa,Herrero-Valea:2020fqa,DelPorro:2023lbv}.   
As pointed out already in \cite{blas_horava_2011}, a singularity at
the universal horizon can dramatically alter the conclusions by
modifying the boundary conditions for the Hawking modes. It will be
interesting to examine if a weak singularity of the type found in this
paper will have such effect.

We have used the obtained solutions to find the black hole sensitivity
$\sigma$ characterizing the dependence of the black
hole inertial mass on its velocity with respect to khronon. To this
end, we devised a simple matching procedure involving only the
khronon perturbations, without the need to compute their back-reaction
on the metric. The sensitivity is independent of the black hole mass
and for phenomenologically acceptable khronon velocity $c_\chi\geq 1$ is
accurately described by a simple analytic expression
(\ref{sigmaexp2}). It depends only on the khronon parameters $\a$ and
$\b$ which are strongly constrained by observations. Thus, $\sigma$
can be at most $(\text{a few})\cdot 10^{-6}$. Its independence of the less
constrained parameter $\l$ is connected to the general property that
for asymptotically flat spacetime the khronon EMT 
vanishes exactly in the limit $\a,\b\to 0$, $\l$-fixed.  

Though small, the sensitivity is non-zero and will affect the dynamics
of black holes in binary systems. It is worth asking if
there can be any observable signatures of Lorentz violation in the
waveforms from black hole binaries to be observed by the next
generation of gravitational wave detectors. 
Unfortunately, the independence of $\s$ of the black hole mass precludes dipolar gravitational radiation in binary systems consisting of two non-rotating black holes. Still, such radiation may arise in black hole -- neutron star and black hole -- white dwarf systems. 

It will be very interesting to generalize our study to the case of moving {\it rotating} black holes along the lines of Refs.~\cite{Frolov:2023gsl,Frolov:2024htj}. These references suggest a possibility of non-trivial interplay between the rotation of a black hole and its motion with respect to khronon. This, in turn, can lead to peculiar effects on the binary system dynamics. In particular, if the sensitivities happen to depend on the black hole spin, they will give rise to dipolar gravitational radiation even in pure black hole binaries. We leave exploration of this promising direction for future.

%==========================================================================
\subsection*{Acknowledgments}
%==========================================================================

We thank Niayesh Afshordi, Enrico Barausse, Diego Blas, Cliff Burgess, Valeri Frolov, Mario
Herrero-Valea, Ted Jacobson, Tsutomu Kobayashi, Duncan O'Dell, Jury Radkovski and Thomas
Sotiriou 
for insightful discussions. 
The work is supported by
the 
Natural Sciences and Engineering Research Council (NSERC) of Canada.
Research at Perimeter Institute is supported in part by the Government
of Canada through the Department of Innovation, Science and Economic
Development Canada and by the Province of Ontario through the Ministry
of Colleges and Universities.

%\newpage
\appendix
\renewcommand{\thesubsection}{\Alph{subsection}}
\renewcommand{\thesection}{\Alph{section}}
\renewcommand{\theequation}{\thesection.\arabic{equation}}

\section{Khronon energy-momentum tensor near universal horizon}
\label{stressenergy}

We have seen in Sec.~\ref{sec:moving} that the solution for the khronon
perturbation $\chi$ has a contribution that behaves as a non-integer
power $\chi\propto (\xi_\star-\xi)^{\g_4^{(\star)}}$ at the universal
horizon. This power is positive, see Eq.~(\ref{g4range}), and the
corresponding perturbation of the aether vector is bounded. However,
the khronon EMT (\ref{khEMT}) contains derivatives of the aether, so
one may worry if it diverges. In this Appendix we analyze this issue
and show that the components of the linearized EMT are finite at the
universal horizon. 

We focus only on the problematic non-analytic contribution since the
EMT coming from the analytic part of $\chi$ is manifestly regular. To
avoid cluttered notation we will denote the non-analytic power simply
by $\g$. Then 
using Eq.~(\ref{aepert}) we find the leading behavior of the aether
perturbations: 
\be
\label{duuh}
\delta u_v\propto \epsilon^{\g+1}\cos\theta~,~~~
\delta u^\xi\propto \epsilon^{\g+2}\cos\theta~,~~~
\delta u_\theta\propto \epsilon^{\g+1}\sin\theta~,~~~~~
\epsilon=\xi_\star-\xi\;,
\ee
where we have omitted irrelevant constant coefficients. Note the
stronger suppression of $\delta u^\xi$ at $\epsilon\to 0$. 
We observe that the first derivatives of $\delta u_\mu$ are at most of
order $O(\epsilon^\g)$ and therefore vanish at the universal
horizon. Thus, the only potentially divergent terms in the EMT are
those containing second derivatives of $\delta u_\mu$
w.r.t. $\xi$. Let us analyze them one by one.

The first line in Eq.~(\ref{khEMT}) contains only one such term,
\begin{equation}
\label{dTalpha}
   \delta T^{(\alpha)}_{\mu\nu} ~\ni~ -\bar u_\mu \bar u_\nu 
\bar u_\rho \nabla_\l\delta u^{\l\rho}
~\ni~ -\bar u_\mu \bar u_\nu 
\bar u_v \d_\xi\delta u^{\xi v}\;,
\end{equation}
where we have omitted a constant coefficient and denoted the
background quantities with an overbar. In the second equality we chose
the value of the dummy index $\l=\xi$ to maximize the potential
divergence and then used the anti-symmetry of $\delta u^{\mu\nu}$. Now
recall that $\bar u_v=U$ is $O(\epsilon)$ at the universal
horizon. Since $\d_\xi^2\delta u_\mu$ is at most of order
$O(\epsilon^{\g-1})$, the term (\ref{dTalpha}) vanishes as
$O(\epsilon^\g)$. 

Next we turn to the second line in Eq.~(\ref{khEMT}). Here the
derivatives act on the aether divergence $\vartheta=\nabla_\mu
u^\mu$. Its perturbation can be easily calculated as
\be
\label{dvartheta}
\delta\vartheta=\frac{1}{\sqrt{-g}}\big[
\d_\xi(\sqrt{-g}\,\delta u^\xi)
+\d_\theta(\sqrt{-g}\,\delta u^\theta)\big]=O(\epsilon^{\g+1})\;.
\ee
Hence the derivatives of $\delta\vartheta$ are $O(\epsilon^\g)$ and
vanish at $\epsilon\to 0$.

The third line in Eq.~(\ref{khEMT}) requires a bit more work. Let us
start with the last term,
\be
\label{dTb1}
\delta T_{\mu\nu}^{(\b 1)}=\frac{1}{2}g_{\mu\nu}\nabla_\l\nabla_\rho
(\bar u^\l\,\delta u^\rho+\delta u^\l\,\bar u^\rho)\;.
\ee 
The most singular contribution contains two $\xi$-derivatives, so we
get,
\be
\label{dTb11}
\delta T_{\mu\nu}^{(\b 1)}\ni g_{\mu\nu}
\d_\xi^2(\bar u^\xi\,\delta u^\xi)=O(\epsilon^\g)\;,
\ee 
which vanishes at the universal horizon. In the remaining terms, let
us keep only the contributions with two derivatives acting on the
perturbation,
\begin{equation}
\label{dTb2}
\delta T^{(\beta 2)}_{\mu\nu} \ni
-\frac{1}{2} \bar u_\mu \nabla_\l\nabla_\nu\delta u^\l
-\frac{1}{2} \bar u^\l \nabla_\l\nabla_\mu\delta u_\nu
+\frac{1}{2} \bar u_\mu \nabla_\l\nabla^\l\delta u_\nu
+(\mu\leftrightarrow \nu)\;.
\end{equation}
The first term is finite due to the relation 
\be
\nabla_\l\nabla_\nu\delta
u^\l=\nabla_\nu\delta\vartheta+R_{\nu\l}\,\delta u^\l\;.
\ee
The rest is analyzed separately for each component. Keeping the most
dangerous terms at each step of the calculation we obtain,
\bseq
\label{dTb2comp}
\begin{align}
\delta T^{(\b 2)}_{vv}\;&\ni\; \bar u_v\nabla_\l\nabla^\l\delta u_v
\;\ni\; g^{\xi\xi}\bar u_v\d_\xi^2\delta u_v=O(\epsilon^\g)\;,\\
\delta T^{(\b 2)}_{v\xi}\;&\ni\; -\frac{1}{2}\bar
u^\l\nabla_\l\nabla_\xi\delta u_v
+\frac{1}{2}\bar u_v\nabla_\l\nabla^\l\delta u_\xi
+\frac{1}{2}\bar u_\xi\nabla_\l\nabla^\l\delta u_v\notag\\
\;&\ni\; -\frac{1}{2}\bar
u^\xi\d_\xi^2\delta u_v
+\frac{1}{2}g^{\xi\xi}\bar u_v \d_\xi^2\delta u_\xi
+\frac{1}{2}g^{\xi\xi}\bar u_\xi \d_\xi^2 \delta u_v\notag\\
\;&=\;-\frac{1}{2}g^{\xi v}\bar u_v\d_\xi^2\delta u_v
+\frac{1}{2}g^{\xi\xi}\bar u_v \d_\xi^2\delta
u_\xi=O(\epsilon^\g)\;,\\
\delta T^{(\b 2)}_{\xi\xi}\;&\ni\;-\bar u^\l\nabla_\l\nabla_\xi \delta
u_\xi
+\bar u_\xi\nabla_\l\nabla^\l\delta u_\xi\notag\\
\;&\ni\;
-\bar u^\xi\d_\xi^2 \delta u_\xi
+g^{\xi\xi}\bar u_\xi\d_\xi^2\delta u_\xi
=-g^{\xi v}\bar u_v\d_\xi^2\delta u_\xi=O(\epsilon^\g)\;,\\
\delta T^{(\b 2)}_{v\theta}\;&\ni\;\frac{1}{2}\bar
u_v\nabla_\l\nabla^\l\delta u_\theta
\;\ni\;\frac{1}{2}g^{\xi\xi}\bar u_v\d_\xi^2\delta u_\theta=O(\epsilon^\g)\;,\\
\delta T^{(\b 2)}_{\xi\theta}\;&\ni\;-\frac{1}{2}\bar
u^\l\nabla_\l\nabla_\xi\delta u_\theta
+\frac{1}{2}\bar u_\xi\nabla_\l\nabla^\l\delta u_\theta\notag\\
\;&\ni\;-\frac{1}{2}\bar
u^\xi\d_\xi^2\delta u_\theta
+\frac{1}{2}g^{\xi\xi}\bar u_\xi\d_\xi^2\delta u_\theta
=-\frac{1}{2}g^{\xi v}\bar u_v\d_\xi^2\delta
u_\theta=O(\epsilon^\g)\;,\\
\delta T^{(\b 2)}_{\theta\theta}&=O(\epsilon^{\g+1})~,~~~~~
\delta T^{(\b 2)}_{\psi\psi}=0\;.
\end{align}
\eseq
We conclude that the non-analyticity of the khronon does not lead to
any divergences in its EMT and in fact 
the corresponding 
contributions vanish at the universal horizon. 
Of course, the total EMT containing also the analytic part need not be
zero at the universal horizon.

\section{Numerical method}
\label{combiningChi}

Here we outline our approach to numerical solution of the khronon
perturbation equation (\ref{perturbationEOM}). This is a linear
fourth order ordinary differential equation with three singular points
$\xi=0,\xi_c,\xi_\star$. Its coefficient functions $A(\xi)$, $B(\xi)$,
$C(\xi)$ are expressed through the background aether configuration,
see Eqs.~(\ref{ABC}), which itself satisfies Eq.~(\ref{ude}) and is
known only numerically. We seek a solution for $\chi$ in the
interval $0<\xi<\xi_\star$ which behaves as $\xi^{-1}$ and
$(\xi-\xi_\star)^{-1}$ at the two boundaries and is regular at
$\xi=\xi_c$. To this aim, we perform the following sequence of steps:\footnote{
We use {\it Mathematica} \cite{Mathematica} and the {\it xAct} package \cite{xAct,Mart_n_Garc_a_2008,Brizuela_2009,Nutma_2014} for the 
analytic and numerical manipulations. }

\paragraph{Step 1.} To avoid singular behavior at the boundaries, we
make a change of variable
\be
\label{tildechidef}
\tilde \chi(\xi)=\xi(\xi-\xi_\star)\chi(\xi)\;.
\ee
This obeys a differential equation
\be
\label{tildechieq}
\sum_{n=0}^4\tilde A_n\,\tilde\chi^{(n)}=0\;,
\ee
where $\tilde\chi^{(n)}$ stands for the $n$th derivative and the
coefficients $\tilde A_n(\xi)$ are analytically expressed through $A$, $B$
and $C$ and hence through the background fields $U$, $V$ and their
derivatives. 

\paragraph{Step 2.} We analytically substitute $V$ and its derivatives
in the expressions for $\tilde A_n$ using the unit-norm relation
(\ref{eqn:unit_norm}). Then we further substitute the second and higher
derivatives of $U$ from the background equation (\ref{ude}) and its
derivatives. In this way we obtain the coefficient functions $\tilde
A_n$ expressed only through $\xi$, $U(\xi)$ and $U'(\xi)$.

\paragraph{Step 3.} We Taylor expand $\tilde A_n$ around the point
$\xi_c$ and determine the solution $\tilde\chi$ through order
$O\big((\xi-\xi_c)^5\big)$ in terms of its value
$a_1\equiv\tilde\chi(\xi_c)$ and the first and second derivatives,
$a_2\equiv\tilde\chi'(\xi_c)$, $a_3\equiv\tilde\chi''(\xi_c)/2$, at this
point. Note that the third derivative of $\tilde\chi$ at $\xi_c$ is
not independent according to the analysis of
Sec.~\ref{sec:moving}. Thus, denoting $\epsilon=(\xi-\xi_c)$ we get,
\begin{equation}
\label{xicexpan}
   \tilde \chi = 
	a_1 + a_2 \epsilon + a_3\epsilon^2 
+  h_1^{(c)}(a_1, a_2, a_3) \epsilon^3 +
h_2^{(c)}(a_1,a_2,a_3)\epsilon^4
+h_3^{(c)}(a_1,a_2,a_3)\epsilon^5 + O(\epsilon^6)\;,
\end{equation}
where $h_1^{(c)}(a_1,a_2,a_3)$ etc. are linear forms with known
numerical coefficients.

Similarly, we get the solution in the form of the power expansions
around the points $\xi=0$ and $\xi=\xi_\star$. Around the point
$\xi_\star$ we have
\begin{equation}
\label{xistarexpan}
   \tilde \chi = 
	a_4 + a_5 \epsilon + h_1^{(\star)}(a_4,a_5)\epsilon^2 +\ldots
+ a_6\,\epsilon^{\g_{4}^{(\star)}}\big(1+g_2^{(\star)} \epsilon +
\ldots\big)+O(\epsilon^6)\;,~~~~ \epsilon=\xi_\star-\xi\;,
\end{equation}
where $\g_4^{(\star)}$ is given in Eq.~(\ref{powerUH}) and $a_4$,
$a_5$, $a_6$ are free parameters. The rest of the coefficients are
linearly 
expressed through them. The expansion
around $\xi=0$ is a bit trickier since it includes also logarithmic
terms which arise generally when there is an integer difference in the powers
of independent solutions \cite{bender}, 
\begin{align}
\label{xi0expan}
    \tilde \chi = 
    &a_7 + a_8\xi + h_1^{(0)}(a_7,a_8) \xi^2 + a_9 \xi^3 + 
h_2^{(0)}(a_7,a_8,a_9)  \xi^4+\ldots \nonumber\\
    &+ (\log\xi) \big(g_1^{(0)}(a_7,a_8)\xi^3 + g_2^{(0)}(a_7,a_8) \xi^4 +
    \ldots\big)
+O(\xi^6)\;,
\end{align}
with $a_7$, $a_8$, $a_9$ being free parameters.

\paragraph{Step 4.} Next we construct a basis of $9$ numerical
solutions. To this end, we step away by $\epsilon\sim 10^{-3}$ from
the singular points and set up initial conditions for numerical
integration of Eq.~(\ref{perturbationEOM}) using the expansions
(\ref{xicexpan}) --- (\ref{xi0expan}). Notice that having expansion
through order $\epsilon^5$ ensures that the third derivative entering 
the initial conditions is determined with second-order
precision in $\epsilon$.

In more detail, we emit three solutions from the point $\xi_c$,
\be
\label{xicsols}
\tilde\chi^{(c)}_{(100)}(\xi)~,~~~~~
\tilde\chi^{(c)}_{(010)}(\xi)~,~~~~~
\tilde\chi^{(c)}_{(001)}(\xi)\;,
\ee 
corresponding to the choices
\be
(a_1,a_2,a_3)=(1,0,0)~,~~(0,1,0)~,~~(0,0,1)
\ee
in the general expression (\ref{xicexpan}). They are defined in the
intervals $[\epsilon,\xi_c-\epsilon]$ and
$[\xi_c+\epsilon,\xi_\star-\epsilon]$. Similarly, the solutions
emitted from $\xi_\star$ and $\xi=0$ are 
\bseq
\begin{align}
\label{xistarsols}
\tilde\chi^{(\star)}_{(100)}(\xi)~,~~~~~
\tilde\chi^{(\star)}_{(010)}(\xi)~,~~~~~
\tilde\chi^{(\star)}_{(001)}(\xi)\;,\\
\label{xi0sols}
\tilde\chi^{(0)}_{(100)}(\xi)~,~~~~~
\tilde\chi^{(0)}_{(010)}(\xi)~,~~~~~
\tilde\chi^{(0)}_{(001)}(\xi)\;,
\end{align}
\eseq
and correspond to the sets
\bseq
\begin{align}
(a_4,a_5,a_6)=(1,0,0)~,~~(0,1,0)~,~~(0,0,1)\;,\\
(a_7,a_8,a_9)=(1,0,0)~,~~(0,1,0)~,~~(0,0,1)\;,
\end{align}
\eseq
in Eqs.~(\ref{xistarexpan}), (\ref{xi0expan}), respectively. The solutions
(\ref{xistarsols}) are defined at
$\xi\in[\xi_c+\epsilon,\xi_\star-\epsilon]$, whereas the solutions
(\ref{xi0sols}) at $\xi\in [\epsilon,\xi_c-\epsilon]$.

\paragraph{Step 5.} 
We now consider linear combinations 
\bseq
\label{lincombs}
\begin{align}
\tilde\chi^{(c)}=a_1\tilde\chi^{(c)}_{(100)}+a_2\tilde\chi^{(c)}_{(010)}
+a_3\tilde\chi^{(c)}_{(001)}\;,\\
\tilde\chi^{(\star)}=a_4\tilde\chi^{(\star)}_{(100)}+a_5\tilde\chi^{(\star)}_{(010)}
+a_6\tilde\chi^{(\star)}_{(001)}\;,\\
\tilde\chi^{(0)}=a_7\tilde\chi^{(0)}_{(100)}+a_8\tilde\chi^{(0)}_{(010)}
+a_9\tilde\chi^{(c)}_{(001)}\;,
\end{align}
\eseq
and require that they describe one and the same solution. To this end,
we match the solutions $\tilde\chi^{(c)}$ and $\tilde\chi^{(\star)}$
through their third derivatives at the midpoint of the interval
$(\xi_c,\xi_\star)$. This produces four linear equations for the
coefficients $a_1,\ldots,a_6$. Four more equations relating $a_1$,
$a_2$, $a_3$ and $a_7$,
$a_8$, $a_9$ are obtained by matching $\tilde\chi^{(c)}$ to
$\tilde\chi^{(0)}$ through third derivatives at the midpoint of
$(0,\xi_c)$. Thus, we have a system of $8$ linear equations for $9$
unknowns $a_1,\ldots,a_9$. Solving it, we express all coefficients
through a single parameter $a_7$ which we set to $1$ according to the
normalization condition (\ref{chi0bc}). This fixes the solution
$\tilde\chi$ completely. \\

\begin{figure}
\begin{center}
\includegraphics[scale=0.6]{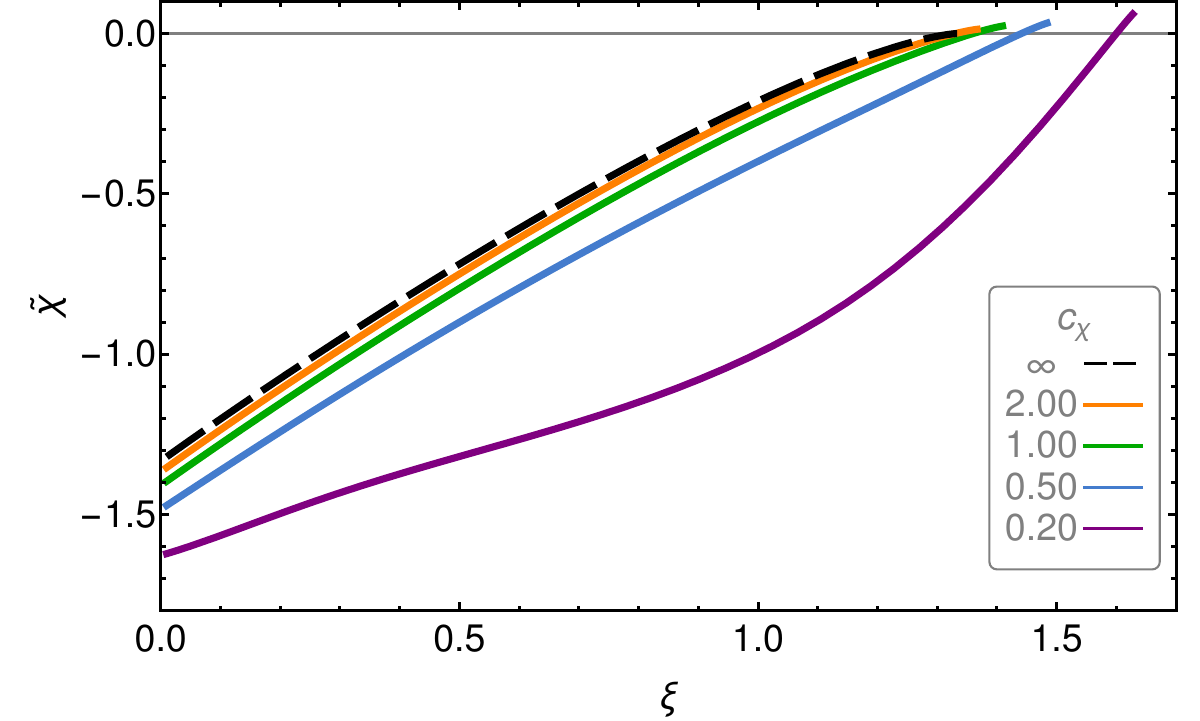}
\caption{Numerical solutions for the function $\tilde\chi$ defined in
  Eq.~(\ref{tildechidef}) for several values of the khronon
  speed.} 
\label{ChiTildediffCx}
\end{center}
\end{figure}

The obtained solutions are displayed in
Fig.~\ref{ChiTildediffCx}. Dividing them by $\xi(\xi-\xi_\star)$ we
get the solutions for the khronon perturbation $\chi$ shown in the
main text.  

\section{Zero modes of $\tilde\Delta$}
\label{app:Lapl}

In Sec.~\ref{sec:infcchi} we encountered an operator $\tilde\Delta$
defined by Eq.~(\ref{tildeDel}). Here we discuss the solutions of the
equation 
\be
\label{tDeq}
\tilde\Delta Y=0
\ee
for a scalar function $Y$. Consider the integral over a spacetime
slice bounded by two leaves of the khronon foliation:
\be
\label{Idef}
I=-\!\!\!\!\!\!\int\limits_{\vf_1\leq\vf\leq\vf_2}\!\!\!\!\!\! 
\dd^4 x\sqrt{-g}\; Y\tilde\Delta Y\;.
\ee
Due to (\ref{tDeq}) we have $I=0$. On the other hand, integrating by
parts we obtain,
\be
\label{I1}
I=\int \dd^4 x\sqrt{-g}\;X^{-1/2} P^{\mu\nu} \nabla_\mu Y\nabla_\nu Y
-\oint \dd\sigma_\mu \sqrt{-g}\;X^{-1/2}P^{\mu\nu} Y\nabla_\nu Y\;.
\ee
The boundary integrals over constant-khronon leaves vanish because on
them $\dd\sigma_\mu P^{\mu\nu}\propto u_\mu P^{\mu\nu}=0$. Next, if the
khronon leaves are topologically trivial and the function $Y$ is
bounded at infinity, the integral over the spatial boundary vanishes
as well. Since $P^{\mu\nu}$ represents a non-degenerate Euclidean
metric on the constant-khronon leaves, the vanishing of $I$ implies
that the gradient of $Y$ along the leaves vanishes. This, in turn,
means that $Y$ must be a function of khronon,
\be
\label{Ysol}
Y(x)=Y\big(\vf(x)\big)\;.
\ee 
In the context of static solutions we obtain $Y={\rm const}$.

In applying the above result to the black hole case we need to deal
with the fact that for a static black hole the khronon foliation
is not topologically trivial due to the presence of the universal
horizon. The problem is overcome by considering the black hole as the
end-point of a dynamical spherical collapse starting from a
topologically trivial foliation. At all finite moments of time the
topology remains trivial and the expression (\ref{Ysol})
applies. Then, by continuity, it also applies to the black hole
obtained in the limit of infinite time.

\section{Expanding khronon in $1/c_\chi^2$}
\label{app:chi0corr}

In Sec.~\ref{sec:infcchi} we considered the full non-linear khronon
equation in the limit $c_\chi=\infty$ and factored out from it the
projected Laplacian $\tilde\Delta$. This allowed us to
obtain a second order differential equation for the khronon
perturbation (\ref{cinfchieq}). It is illuminating to see how this
simplification happens directly at the level of the khronon
perturbation equation (\ref{perturbationEOM}) valid for general values of $c_\chi$. 
This will also give us
access to the $1/c_\chi^2$ terms in the khronon solution. 

Dividing the coefficient functions (\ref{ABC}) by $c_\chi^2$ and taking
$c_\chi\to\infty$ we obtain their limiting form which we denote as
$A_\infty$, $B_\infty$, $C_\infty$. Then it is straightforward to
verify that the differential operator on the left-hand side of
Eq.~(\ref{perturbationEOM}) becomes a total square,
\be
\label{Dopcinf}
(A_\infty\chi'')''-(B_\infty\chi')'+C_\infty\chi=
-\frac{1}{\xi^4}\big(-\xi^4\d_\xi( U_\infty^3\d_\xi)+2\xi^2 U_\infty\big)^2\chi\;,
\ee
where the background function $U_{\infty}$, as well as $V_\infty$
entering in the construction of $A_\infty$, $B_\infty$, $C_\infty$ are
taken from Eq.~(\ref{cinfbkg}). The operator in brackets is
precisely the projected Laplacian (\ref{tildeDel}) 
restricted to the dipole sector,
\be
\label{DLapll1}
\tilde\Delta_{\rm d}\, y(\xi)=-\xi^4(U^3 y')'+2\xi^2 Uy\;.
\ee 
The equation $\tilde\Delta_{\rm d}\,y=0$ does not have any non-trivial bounded
solutions. Hence, we can ``cancel'' one factor $\tilde\Delta_{\rm d}$
and arrive at the equation 
\be
\label{Deqcinf}
-\xi^4(U_\infty^3 \chi')'+2\xi^2 U_\infty \chi=0\;,
\ee
which coincides with (\ref{cinfchieq}), up to an overall
multiplicative factor. The solution considered in
Sec.~\ref{sec:infcchi} vanishes at the universal horizon
$\xi_\star=4/3$ but diverges at $\xi=0$ as $1/\xi$. Let us denote it by
$\chi_\infty(\xi)$. Below we will need the second linearly independent
solution of Eq.~(\ref{Deqcinf}) which vanishes at $\xi=0$ and diverges
at $\xi=4/3$. We denote the second solution by
$\hat\chi_\infty(\xi)$. Its Wronskian with $\chi_\infty(\xi)$ is
proportional to $U_\infty^{-3}(\xi)$; we normalize 
$\hat\chi_\infty(\xi)$ by setting the proportionality coefficient to $1$:
\be
\label{DWron}
\hat\chi'_\infty\chi_\infty-\hat\chi_\infty\chi'_\infty=\frac{1}{U_{\infty}^3(\xi)}\;.  
\ee
This implies that at $\xi\to 0$ the second solution behaves as 
\be
\label{Dhatchiexp}
\hat\chi_\infty=\frac{\xi^2}{3}+O(\xi^3)\;.
\ee

At general values of $c_\chi$ the factorization of the operator in
Eq.~(\ref{perturbationEOM}) is no longer possible. Consider the
combination
\be
\label{DDdef}
\hat
D\chi\equiv (A\chi'')''-(B\chi')'+C\chi+\frac{c_\chi^2}{\xi^4}\tilde\Delta_{\rm
  d}^2\,\chi\;, 
\ee
where $\hat D$ is understood as a differential operator. Note that in
the Laplacian here we use the background function $U$ corresponding to
the finite value of $c_\chi$. Due to the equality (\ref{Dopcinf}), the
operator $\hat D$ remains finite when $c_\chi\to\infty$. It contains
two types of contributions. One of them is due to the parts of the
coefficients $A$, $B$, $C$ without the explicit $c_\chi^2$ factors,
see Eqs.~(\ref{ABC}). Denoting these parts by $A_1$ etc., we have
\be
\label{DD1}
\hat D_1\chi=(A_1\chi'')''-(B_1\chi')'+C_1\chi\;.
\ee  
Second contribution comes from the mismatch between the parts of 
$A$, $B$, $C$ explicitly multiplied by $c_\chi^2$ and the coefficients
coming from the square of the projected Laplacian. The mismatch
appears only in the coefficients of the second and lower
derivatives. A straightforward calculation yields
\be
\label{DD2}
\hat D_2\chi=D_{2,2}(\xi)\chi''+D_{2,1}(\xi)\chi'+D_{2,0}(\xi)\chi\;,
\ee
with 
\bseq
\label{DD2coeffs}
\begin{align}
D_{2,2}=&c_\chi^2\big(6\xi^2U^4V^2-3\xi^4U^4VV''\big)\;,\\
D_{2,1}=&c_\chi^2\big(12\xi
U^4V^2+24\xi^2U^3U'V^2+12\xi^2U^4VV'-12\xi^3U^4VV''\notag\\
&~~~~ -12\xi^4U^3U'VV''-3\xi^4U^4V'V''-3\xi^4U^4VV'''\big)\;,\\
D_{2,0}=&c_\chi^2\big(-4U^2V^2+2\xi^2U^2VV''\big)\;.
\end{align}
\eseq
We now use the background equation in the form (\ref{khreq_bkg1}) to
get rid of the higher derivatives of $V$. This leads to the
cancellation of the $c_\chi^2$ factors and we end up with
\bseq
\label{DD2coeffs1}
\begin{align}
D_{2,2}=&-3\xi^4U^3U''V^2\;,\\
D_{2,1}=&-12\xi^3U^3U''V^2-9\xi^4U^2U'U''V^2-6\xi^4U^3U''VV'
-3\xi^4U^3U'''V^2\;,\\
D_{2,0}=&2\xi^2UU''V^2\;.
\end{align}
\eseq

Above considerations imply that Eq.~(\ref{perturbationEOM}) is
equivalent to
\be
\label{Deqpert}
\frac{1}{\xi^4}\tilde\Delta_{\rm d}^2\,\chi=\frac{1}{c_\chi^2}\hat D\chi\;,
\ee
where $\hat D$ is given by the sum of Eqs.~(\ref{DD1}) and
(\ref{DD2}). The new form is convenient for a perturbative solution in
the inverse powers of $c_\chi^2$. Let us focus on the first correction
to the $c_\chi=\infty$ solution. Of particular interest for us is its
behavior in the vicinity of $\xi=0$ which provides the black hole
sensitivity, as discussed in Sec.~\ref{sec:sens}. 

Our strategy is to solve Eq.~(\ref{Deqpert}) in two steps stripping
off the two factors of Laplacian one after another. Introducing
$y\equiv \tilde\Delta_{\rm d} \chi$ we obtain
a Laplace-type equation with non-vanishing source, 
\be
\label{DLalpleq1}
\frac{1}{\xi^4}\tilde\Delta_{\rm d}\,y=\frac{1}{c_\chi^2}\hat
D\chi_\infty\;, 
\ee 
where on the right-hand side we can evaluate all
functions in the $c_\chi=\infty$ limit.
We construct the solution using the Green's function of the operator
$\xi^{-4}\tilde\Delta_{\rm d}$: 
\be
\label{DyGreen}
y(\xi)=\frac{1}{c_\chi^2}\int \dd\xi'\,{\cal G}(\xi,\xi')\,\hat
D\chi_\infty(\xi')\;. 
\ee
The latter can again be computed for $c_\chi=\infty$ using the two
independent solutions of the homogeneous equation discussed above,
\be
\label{Greens}
{\cal
  G}(\xi,\xi')=\Theta(\xi-\xi')\chi_\infty(\xi)\hat\chi_\infty(\xi')
+\Theta(\xi'-\xi)\hat\chi_\infty(\xi)\chi_\infty(\xi')\;,
\ee
where $\Theta$ is the Heaviside function. Only the second term
contributes into $y(\xi)$ at small $\xi$. Using
Eq.~(\ref{Dhatchiexp}), we get 
\be
y(\xi)=\frac{\xi^2}{3c_\chi^2}\int_0^{4/3}\dd\xi'\,\chi_{\infty}(\xi')
\hat D\chi_\infty(\xi')+O(\xi^3)\;.
\ee 
It remains to solve the equation 
\be
\label{Dlasteq}
\tilde \Delta_{\rm d}\, \chi(\xi)=y(\xi)
\ee
to completely determine the correction. We do not need to do
it, however, if our goal is only to get the sensitivity. It 
suffices to substitute the expansion (\ref{chilead}) into
Eq.~(\ref{Dlasteq}) and require that the terms of order $\xi^2$ on
both sides of the equation match. This gives us
\be
\label{Dchi0int}
\chi_0=-\frac{1}{4}+\frac{1}{6c_\chi^2}
\int_0^{4/3}\dd\xi'\,\chi_{\infty}(\xi')
\hat D\chi_\infty(\xi')\;.
\ee
Numerically evaluating the integral we arrive at Eq.~(\ref{kappa}) from the
main text.

%\newpage
%\bibliographystyle{plain} % We choose the "plain" reference style
\bibliographystyle{JHEP}
\bibliography{references} 
 
\end{document}